\documentclass[showpacs,aps,prd,floatfix,amsmath,amssymb,nofootinbib]{revtex4-2}
\usepackage[utf8x]{inputenc}
\usepackage{graphicx}
\usepackage{amsmath}
\usepackage{array}
\usepackage{slashed}	
\usepackage[makeroom]{cancel}
\usepackage{bbding}
\begin{document}
\title{Rare decay $t\to c\gamma\gamma$   via  scalar leptoquark doublets}
\author{A. Bola\~nos--Carrera}
\affiliation{Tecnol\'ogico de Monterrey, Department of Science, Campus Puebla, Av. Atlixc\'ayotl 2301, CP 72453, Puebla, Puebla, M\'exico}
\author{G. Tavares--Velasco}
\affiliation{Facultad de
Ciencias F\'\i sico Matem\'aticas, Benem\'erita Universidad
Aut\'onoma de Puebla, Apartado Postal 1152, Puebla, Pue., M\'
exico}
\author{R. S\'anchez--V\'elez}
\email[Corresponding author: ]{ricsv05@icloud.com}
\affiliation{ Departamento de F\'isica, Centro de Investigaci\'on y de Estudios Avanzados del IPN
Apdo. Postal 14-740 07000 Ciudad de M\'exico, M\'exico}

\begin{abstract}
A calculation of the one-loop contribution to the rare three-body flavor changing neutral current top quark decay $t\to c\gamma\gamma$ is presented in the framework of models with one or more scalar leptoquark (LQ) $SU(2)$ doublets with hypercharge $7/6$. Analytical expressions for the invariant amplitude of the generic decay $f_i\to f_j\gamma\gamma$, with $f_{i,j}$ a lepton or quark, are presented in terms of Passarino-Veltman integral coefficients, from which the amplitudes for the processes $t\to c\gamma\gamma$ and $\ell_i\to \ell_j\gamma\gamma$ follow easily. An analysis of the current constraints on the parameter space is presented in the scenario with only one scalar LQ doublet and   bounds on the LQ couplings are obtained  from the  muon $g-2$ anomaly, the lepton flavor violating (LFV) decay $\tau\to \mu\gamma$ and  extra constraints meant to avoid tension between theory predictions and experimental data. For a LQ with a mass in the range of $1$--$1.5$ TeVs, the estimate ${\rm Br}(t\to c\gamma\gamma)\sim 10^{-11}$--$10^{-12}$ is obtained for the largest allowed values of the LQ coupling constants, which means that this decay would be below the reach of future experimental measurements.
We also consider an scenario with three scalar doublets, which was recently proposed to explain the lepton flavor universality violation anomalies in $B$ decays as well as the muon $g-2$ anomaly. Although this scenario allows large LQ couplings to the tau lepton and the $c$ and $t$ quarks, the
branching ratio of the $t\to c\gamma\gamma$ decay is also of the order of $10^{-11}$--$10^{-12}$ for LQ masses o the order of f 1.7 TeV.
\end{abstract}
\maketitle

\section{Introduction}
\label{introduction}
The study of flavor changing neutral currents (FCNCs) has long been a topic of great interest both theoretically and experimentally \cite{Chakraborty:2003iw,Larios:2006pb}. This class of effects is considerably suppressed according to the standard model (SM), where they arise up to the one-loop level \cite{Eilam:1990zc,DiazCruz:1989ub}. Therefore FCNC transitions could provide signals of new physics and shed light on any possible SM extension. In the experimental side, the advent of the large hadron collider (LHC) offers a great potential to  search for signals of various rare FCNC top quark  decays, such as $t\to c V$ ($V=\gamma,g, Z$), $t\to cH$, $t\to c\ell^-\ell^+$,   $t\to c \gamma\gamma$, $t\to c gg$, $t\to c\gamma H$, and $t\to c \gamma Z$.    While the two-body decays  $t\to c V$ and $t\to cH$ have been largely studied in the context of the SM and several of its extensions \cite{Eilam:1990zc,Lu:1996ji,Wang:1994qd,Lu:1998gm,Lu:2003yr,Couture:1994rr,Li:1993mg,Lopez:1997xv,Yang:1997dk,Frank:2005vd,GonzalezSprinberg:2007zz,CorderoCid:2005kp,CorderoCid:2004vi,Cortes-Maldonado:2013rca}, less attention has been paid to the three-body decays as they are expected to be more suppressed,  but also because they involve  lengthy and cumbersome  calculations.

The decay $t\to c\gamma$ can only arise at the one-loop level due to electromagnetic gauge invariance and  is further suppressed due to the Glashow-Iliopoulos-Maiani (GIM) mechanism, so its branching ratio is considerably small, of the
order of $10^{-10}$ \cite{Eilam:1990zc,DiazCruz:1989ub}. However, in other SM extensions such a decay may not be  GIM-suppressed and its branching ratio $Br(t\to c\gamma)$ can be enhanced  by several orders of magnitude, ranging from values of the order of $ 10^{-7}$ in   two-Higgs doublet models \cite{Eilam:1990zc} up to $10^{-5}$ in supersymmetric models  \cite{Couture:1994rr,Li:1993mg,Lopez:1997xv,Yang:1997dk}. Furthermore, some time ago, it was pointed out that there are some new physics scenarios where the three-body decay $t\to c\gamma\gamma$ could have  a larger branching ratio than those of the two-body decays $t\to c\gamma$ \cite{Diaz-Cruz:1999wcs}. Thus, it is worth studying the rare three-body FCNC top quark decays in extension models despite the complexity involved in the respective calculation in order to assess if they could be at the reach of experimental detection. Along this line, the  decay $t\to c\gamma\gamma$ has been studied in the framework of the little Higgs model with T-parity \cite{Han:2011xd,Han:2016hef} and also in a top-color assisted technicolor theory \cite{Yue:2001cy}.

In this work we will present a calculation of  the $t\to c\gamma\gamma$  decay in the framework of leptoquark (LQ) models.
Two-body FCNC top quark decays have already been calculated in the context of this class of models: a calculation of the contribution of a model with an $SU(2)$ scalar LQ doublet  to the two-body decays $t\to cX$ ($X=\gamma, Z, H, g$) and also to the three-body decay $t\to c\ell^-\ell^+$ was presented  in Ref. \cite{Bolanos:2019dso} along with a comprehensive analysis of the parameter space of the model consistent with the then current constraints from direct LQ searches at the LHC, the Higgs boson coupling modifiers, the muon $g-2$ anomaly, and the lepton flavor violating  decay (LFV) $\tau\to \mu \gamma$. To our knowledge there is no previous calculation of the contribution of LQs to the  three-body FCNC top quark rare decay $t\to c\gamma\gamma$.

LQs  are  hypothetical particles carrying both lepton and color number that  were proposed long ago in the Pati-Salam model \cite{Pati:1974yy} and also in the context of Grand Unification theories \cite{Georgi:1974sy,Georgi:1974yf,Fritzsch:1974nn,Dimopoulos:1980hn,Senjanovic:1982ex,Frampton:1989fu}, though they can also arise naturally in theories with composite fermions \cite{Schrempp:1984nj,Buchmuller:1985nn,Gripaios:2009dq}, superstring-inspired $E_6$ models \cite{Witten:1985xc,Hewett:1988xc}, technicolor models \cite{Ellis:1980hz,Farhi:1980xs,Hill:2002ap}, etc.  A shortcoming of some of these models is that  they could  allow dangerous LQ diquark couplings that would induce proton decay at the tree-level \cite{Langacker:1980js}, so additional symmetries must be invoked to preserve proton stability.
In addition, if LQs couple to the first fermion generation,  large contributions  to low-energy observable quantities can arise, such as atomic parity violation, parity-violating electron scattering, coherent neutrino-nucleus scattering, and electroweak precision parameters, which together with direct searches at the LHC set tight constraints  on the parameter space of such LQs \cite{Davidson:1993qk,Shanker:1982nd,Shanker:1981mj,Leurer:1993qx,Leurer:1993em,Crivellin:2021egp}. However, LQ couplings to the second and third fermion families are not strongly constrained yet and  recently LQ particles have become the source of renewed  attention in the literature (for a recent LQ review see \cite{Dorsner:2016wpm}) since they could explain the  lepton flavor universality  violating (LFUV) effects hinted at  semi-leptonic $B$ decays and can also provide a solution to the muon $g-2$ anomaly \cite{Sakaki:2013bfa,Becirevic:2017jtw,Crivellin:2017zlb,Becirevic:2018afm,Angelescu:2018tyl,Buttazzo:2017ixm,Marzocca:2021azj,Angelescu:2021lln,Bauer:2015knc,Becirevic:2016yqi,Kumar:2018kmr,Cheung:2022zsb,Bhaskar:2022vgk,Datta:2019bzu,Hati:2019ufv,DaRold:2019fiw,Cornella:2019hct,Mandal:2018kau,Alonso:2015sja,Calibbi:2015kma,Hiller:2016kry,Bhattacharya:2016mcc,Barbieri:2015yvd,Barbieri:2016las,Calibbi:2017qbu,Crivellin:2017dsk,Bordone:2018nbg,Crivellin:2018yvo,Bordone:2019uzc,Bernigaud:2019bfy,Aebischer:2018acj,Fuentes-Martin:2019ign,Popov:2019tyc,Fajfer:2015ycq,Blanke:2018sro,deMedeirosVarzielas:2019lgb,deMedeirosVarzielas:2015yxm,Crivellin:2019dwb,Saad:2020ihm,Saad:2020ucl,DaRold:2020bib,Bordone:2017bld,Biswas:2018snp,Heeck:2018ntp,Sahoo:2015wya,Chen:2016dip,Dey:2017ede,Chauhan:2017ndd,Fajfer:2012jt,Freytsis:2015qca,Li:2016vvp,Zhu:2016xdg,Popov:2016fzr,Deshpande:2016yrv,Becirevic:2016oho,Cai:2017wry,Altmannshofer:2017poe,Kamali:2018fhr,Azatov:2018knx,Kim:2018oih,Aydemir:2019ynb,Yan:2019hpm,Marzocca:2018wcf,Bigaran:2019bqv,BhupalDev:2020zcy,Altmannshofer:2020axr,Fuentes-Martin:2020bnh,Gherardi:2020qhc,Chakraverty:2001yg,Cheung:2001ip,Biggio:2016wyy,ColuccioLeskow:2016dox,Chen:2017hir,Das:2016vkr,Crivellin:2018qmi,Kowalska:2018ulj,Dorsner:2019itg,Bigaran:2020jil,Dorsner:2020aaz,Babu:2020hun,Crivellin:2020tsz}.
Furthermore, LQs can be accommodated in models where neutrino mass is generated radiatively \cite{AristizabalSierra:2007nf,Saad:2020ihm,Babu:2020hun,Zhang:2021dgl,Cai:2017wry,Popov:2016fzr,Babu:2020hun}. It is thus worth calculating LQ contributions to FCNC rare top quark decays.

The rest of the presentation is organized as follows. In Section \ref{model} we present an overview
of LQ models and focus on a minimal renormalizable scalar LQ model with no proton decay, where there are potential sources of flavor change in the quark sector induced by  scalar LQs. Section \ref{calculation} is devoted to discuss
the calculation of the $t\to c\gamma\gamma$  decay amplitude in our LQ model: for the sake of completeness, the invariant amplitude for the general fermion decay $f_i\to f_j\gamma\gamma$ is obtained via the Passarino-Veltman reduction method and the corresponding form factors are  presented in Appendix \ref{FormFactors}. From these expressions the decay width for the  $t\to c\gamma\gamma$ process follows straightforwardly. The numerical analysis of the LQ parameter space consistent with the current experimental constraints, considering two potential scenarios for the LQ couplings along with the numerical evaluation of the $t\to c\gamma\gamma$  branching ratio are presented in Sec. \ref{paramcons}. Finally, Sec. \ref{conclusions} is devoted to the conclusions and outlook.

\section{A renormalizable scalar LQ model with proton stability}
\label{model}
We now present the theoretical framework  required for the calculation of the rare top quark decay $t\to c\gamma\gamma$  focusing on a model where there is no dangerous contribution to proton decay. LQs are hypothetical particles carrying both lepton and color number, thereby coupling simultaneously to lepton and quarks. A systematic classification of all $SU(3)_C\times SU(2)_L\times U(1)_Y$ LQ representations and their renormalizable couplings to the SM fields was presented in Ref. \cite{Buchmuller:1986zs}  via effective Lagrangians. According to all possible representations of the SM gauge group, it was found that LQs can be accommodated  in ten   representations: five scalar ones and five vector ones.

As far as vector LQs are concerned, they arise in grand unification theories and may be troublesome as  can trigger rapid proton decay, which sets a lower constraint of $10^{16}$ GeV on the mass of such gauge LQs \cite{Langacker:1980js,Georgi:1974yf}. Even if an ad-hoc symmetry is imposed to forbid proton decay, the mass of vector LQs can  be strongly constrained by rare $K$, $\pi$ and $B$ meson decays \cite{Valencia:1994cj,Kuznetsov:1994tt}.
However,  there are two phenomenologically viable vector LQ models at the TeV scale with no proton decay, which involve the $(3,1,2/3)$ and $(3,3,2/3)$ vector LQ representations \cite{Assad:2017iib}. Models based on these representations have been studied recently  as they can explain the LFUV anomalies in $B$-meson decays and still be consistent with current experimental constraints from $K$ and $B$ meson decays, electroweak precision observable parameters as well as direct searches at the LHC \cite{Fajfer:2015ycq,Calibbi:2017qbu,Crivellin:2018yvo,Datta:2019bzu,Hati:2019ufv,Hati:2019ufv,DaRold:2019fiw,Cornella:2019hct,Angelescu:2021lln,Bhaskar:2021pml,Cheung:2022zsb}. In particular, the  $(3,1,2/3)$ vector LQ representation $U_1^\mu$ is attractive as it is predicted by  the minimal realization of the Pati-Salam model \cite{Pati:1974yy} and also due to the absence of tree-level  contributions to the very constrained decay $B\to K^*\nu\bar\nu$  \cite{Assad:2017iib,Heeck:2018ntp}.

As for scalar LQ models, there are two renormalizable scalar LQ models \cite{Arnold:2013cva} that do not have proton decay at the tree-level, which involve the $(3,2,7/6)$ and $(3,2,1/6)$ scalar LQ representations.
For the purpose of this work, we are interested in LQ models with the scalar LQ $(3,2,7/6)$ representation, which is usually denoted as $R_2$ in the literature \cite{Buchmuller:1986zs}. This scalar $SU(2)$ LQ doublet  representation   provides  simple renormalizable models that conserve  baryon number \cite{Arnold:2013cva}, thereby forbidding  dangerous   contributions to  proton decay. The phenomenology of LQ models with  scalar doublets $R_2$ has been extensively studied in the literature: for representative works see for instance \cite{Arnold:2013cva,Bolanos:2013tda,Mohanta:2013lsa,Allanach:2015ria,Dey:2015eaa,Sahoo:2015fla,Baek:2015mea,Sahoo:2017lzi,Sheng:2018qtp,Mandal:2019gff,Chandak:2019iwj,Dekens:2018bci,Becirevic:2017jtw, Popov:2019tyc, Angelescu:2021lln, Angelescu:2018tyl, Crivellin:2021egp, Crivellin:2021bkd, He:2021yck, Bigaran:2020jil, Bigaran:2021kmn, deMedeirosVarzielas:2015yxm, Sahoo:2015wya, Kamali:2018fhr, Chakraverty:2001yg, Cheung:2001ip, Queiroz:2014pra, ColuccioLeskow:2016dox, Chen:2017hir, Kowalska:2018ulj,Bansal:2018eha,Bolanos:2019dso, Dorsner:2020aaz, Crivellin:2020tsz, Iguro:2020keo,Husek:2021isa,Becirevic:2022tsj} and
they have been the source of attention recently as  can explain the apparent discrepancies between the SM predictions and experimental measurements. For instance, models with only one $R_2$ scalar LQ doublet can  explain the muon anomaly and the $R_{D,D^*}$ anomalies \cite{Cheung:2001ip,ColuccioLeskow:2016dox,Crivellin:2020tsz}, though the authors of Ref. \cite{Crivellin:2022mff} showed that multiple  $R_2$ scalar LQ doublets are necessary to explain the apparent $R_{K,K^*}$ anomalies \cite{LHCb:2017avl,LHCb:2021trn}, which however seem to be excluded by the most recent measurements of  the $b\to s \ell^+\ell^-$ decay by the LHCb collaboration \cite{LHCb:2022qnv}. In addition, another appealing feature of models with $R_2$ scalar  LQ doublets is that they can provide a mechanism of  neutrino mass generation, such as in the model studied in   \cite{AristizabalSierra:2007nf,Saad:2020ihm,Zhang:2021dgl,Popov:2016fzr}, where neutrinos masses are generated  via the mixing of $R_2 $ with an extra LQ singlet $S_1$ through radiative corrections.

In this work we will consider LQ models where the SM is augmented with one or more  $SU(2)$  scalar doublets $R_2$  that induce the FCNC decay $t\to c\gamma\gamma$.   We will present the theoretical framework for a lone scalar doublet $R_2$ and the extension for models with multiple scalar LQ doublets $R_2$ will follow straightforwardly. As already mentioned, the presence of  additional scalar LQ doublets $R_2$ is meant to explain the LFUV anomalies in $b$-hadron decays, though   additional contributions to FCNC top quark decays are not necessarily expected  since  extra symmetries may be required to meet the current experimental constraints, thereby imposing tight constraints on the parameter space of the model.

The $R_2$ LQ doublet has hypercharge $7/6$, thereby giving rise to two scalar LQs with electric charges $2/3$ and $5/3$, which we denote by $\Omega_{2/3}$ and $\Omega_{5/3}$, where the subscripts stand for the LQ electric charge. Both $\Omega_{2/3}$ and $\Omega_{5/3}$  predict rich phenomenology as already noted and can give new contributions to several observable quantities (for a recent review of LQ constraints from experimental data see \cite{Dorsner:2016wpm}), such as atomic parity violation \cite{Langacker:1990jf,Leurer:1993em}, meson decays \cite{Shanker:1981mj,
Leurer:1993em,Davidson:1993qk,Valencia:1994cj,Dorsner:2016wpm}, electric dipole moments of leptons \cite{Arnold:2013cva},  LFV lepton decays \cite{Lavoura:2003xp,Gabrielli:2000te}, oscillations in $K$, $D$ and $B$ meson systems \cite{Shanker:1982nd,Leurer:1993ap}, oblique corrections \cite{Keith:1997fv,Crivellin:2020ukd}, electroweak precision observable parameters \cite{Mizukoshi:1994zy,deBlas:2014mba}, Higgs boson modifiers \cite{Crivellin:2020ukd,Bolanos:2019dso}, etc., which along with direct searches at the LHC \cite{CMS:2018ncu,ATLAS:2020dsk,ATLAS:2020xov,CMS:2021far,ATLAS:2022wcu} set strong constraints on the corresponding parameter space.
For the purpose of this work we are interested in  the charge $5/3$ LQ  as it couples to left- and right-handed fermions simultaneously and can induce new physics effects in the LFV decays $H\to \mu\tau$ \cite{Baek:2015mea} and $\ell_i\to \ell_j\gamma$  \cite{Lavoura:2003xp,Gabrielli:2000te,Husek:2021isa,Bolanos:2019dso} as well as the FCNC top quark decays $t\to c\gamma$, $t\to cZ$, and $t\to cH$ \cite{Bolanos:2019dso}. This LQ can also give contributions to the  $t\to c\gamma\gamma$  decay, which is the  topic of interest of the present work.

The Yukawa lagrangian for an $R_2$ doublet can be written as
\begin{equation}
\mathcal{L}_{F=0}=Y^{RL}_{ij} {R}_2^T\bar u^{'i}_+{R}i\tau_2 L_L^{'j}+Y^{LR}_{ij}\bar{Q}^{'i}_L{e'}^j_R R_2+\text{H.c}.,
\end{equation}
where as usual $L_L^{'i}$ and $Q_L^{'i}$ are $SU(2)_L$ left-handed lepton and quark doublets, respectively, whereas $e_R^{'i}$ and $q_R^{'i}$ are $SU(2)$ singlets, with $i$ and $j$ being generation indices. For the Yukawa couplings we use the notation of Refs. \cite{Crivellin:2021ejk,Crivellin:2022mff} and consider that the $R_2$ LQ only couples  to the fermions of the second and third generations since the couplings to the fermions of the first generation  are strongly constrained by low-energy data, such as atomic parity violation  \cite{Langacker:1990jf,Leurer:1993em,Davidson:1993qk}, universality in leptonic pion decays \cite{Shanker:1982nd,Davidson:1993qk,Leurer:1993em},
$\mu e$ conversion \cite{Shanker:1981mj},
flavor changing kaon decays \cite{Shanker:1981mj,Valencia:1994cj},
 and $K^0- \overline{K}^0$ and $D^0-\overline{D}^0$ mixing \cite{Shanker:1982nd,Davidson:1993qk,Leurer:1993ap}. For other recent analyses see \cite{Crivellin:2021egp,Bansal:2018eha}.

After electroweak symmetry breaking we rotate the $SU(2)_L$ LQ doublet into its mass eigenstates: $R_2^T= \left(\Omega_{5/3}\;\; \Omega_{2/3}\right)$, which at the lowest order in $\upsilon$ coincide with the weak eigenstates \cite{Crivellin:2020ukd}. As for the weak eigenstates of the up and down quarks, we will consider two scenarios. In the so-called up-aligned scenario, the weak eigenstates of the up quarks $u^{'i}$ are chosen as the mass eigenstates $u^i$, whereas the weak eigenstates of the down quarks $d^{'i}$ are rotated to the mass eigenstates $d^{i}$ via  the Cabibbo-Kobayashi-Maskawa mixing matrix $V_{CKM}$: $d^{'i}\to V^{ik}_{\text{CKM}} d^{k}$. In this scenario the LQ interactions with the SM fermions  read
\begin{equation}
\label{YukawaLag}
\mathcal L_{F=0}=  \left(Y^{RL}_{ij}\bar u^i P_L e^j+Y^{LR}_{ij}\bar u^i P_R e^j\right)\Omega_{5/3}+\hat Y^{LR}_{ij}\bar d^iP_R e^j\Omega_{2/3}-Y^{RL}_{ij}\bar u^iP_L \nu_j \Omega_{2/3}+\text{H.c.},
\end{equation}
where $P_{L,R}$ are the chiral projection operators and we define $\hat Y^{LR}_{ij}=V^{ik}_{\text{CKM}}Y^{LR}_{kj}$.
Alternatively, in the down-aligned scenario one sets $d^{'i}=d^i$ and $u^{'i}= V^{ik}_{\text{CKM}} u^{k}$. This choice only affects the LQ interactions  with left-handed up quarks, so the replacement $\bar u^i P_R e^j\Omega_{5/3}\leftrightarrow \bar d^iP_R e^j\Omega_{2/3}$ must be made in Eq. \eqref{YukawaLag} to obtain the corresponding interaction Lagrangian.

Below we will analyze two LQ models (Scenario I and scenario II) and  constraints on  LQ Yukawa couplings will be obtained from experimental data. In Scenario I we consider a model with only one $R_2$ scalar LQ doublet that is not primarily meant to explain the LFUV anomalies in $B$-meson decays: the $R_2$ LQ doublet could be assumed as a piece of a more complete model provided with another mechanism to explain such anomalies, as long as  they are confirmed by future measurements.
In this scenario we do not assume a particular pattern for the LQ Yukawa couplings, which are then bounded from the  muon $g-2$ anomaly, the LFV decay $\tau\to \mu\gamma$, and other experimental constraints.  As far as scenario II is concerned, we consider the three-LQ-doublet model recently proposed  to address the $B$-meson anomalies \cite{Crivellin:2022mff}, which require that LFV is forbidden.  In such a model, the down-aligned scenario was assumed and a comprehensive analysis of the allowed regions of the parameter space of the model consistent with experimental constraints was performed \cite{Crivellin:2022mff}.
We will not explore other alternative models with a specific structure for the LQ Yukawa couplings here as the two scenarios considered in our analysis will allow us to assess the order of magnitude of the branching ratio for the $t\to c\gamma\gamma$ decay.

As for the LQ couplings to the photon, they can be obtained from the  kinetic  lagrangian
\begin{equation}
\label{LQKinetic}
{\cal L}_{\text{kin}}=\frac{1}{2}(D_\mu R_2)^\dagger D^\mu R_2,
\end{equation}
where the $SU(2)_L\times U(1)_Y$ covariant derivative is  given by
\begin{equation}
D_\mu R_2= \left(\partial_\mu+ig\frac{\tau^i}{2} W^i_\mu+i g'\frac{7}{6}B_\mu\right)R_2.
\end{equation}
Thus, the LQ couplings to one and two photons  can be simply written as
\begin{equation}
\mathcal{L}_{\text{kin}}\supset\sum_{Q=2/3,5/3}\left( iQ_Q A^\mu\left(\Omega_Q^*\partial_\mu \Omega_Q-\Omega_Q\partial_\mu \Omega_Q^* \right)+Q_Q^2A_\mu A^\mu\Omega_Q^*\Omega_Q\right),
\end{equation}
where the subscript $Q$ denotes the LQ electric charge.
Finally, we consider the following renormalizable effective  LQ interactions to the SM Higgs doublet $\Phi$
\begin{equation}\label{HiggsLQ}
\mathcal{L} = \left( M_{R_2}^2+\lambda_{R_2}  \Phi^\dagger \Phi\right)\left(R_2^\dagger R_2 \right),
\end{equation}
where   $M_{R_2}$ is the LQ mass matrix. After rotating to the mass eigenstates the LQ masses become non-degenerate at the lowest order in $\upsilon^2$.  We can also obtain the Higgs boson coupling to the LQs:
\begin{equation}
\label{ScaLagI}
{\mathcal L}\supset \sum_{Q=2/3,5/3} \lambda_{\Omega_{Q}} v H \Omega_Q^* \Omega_Q,
\end{equation}
which can be useful to constrain the LQ couplings to the SM Higgs boson from the Higgs coupling modifiers $\kappa_\gamma$ and $\kappa_{g}$ \cite{Bolanos:2019dso}.

\begin{figure}[hbt!]
  \centering
  \includegraphics[width=10cm]{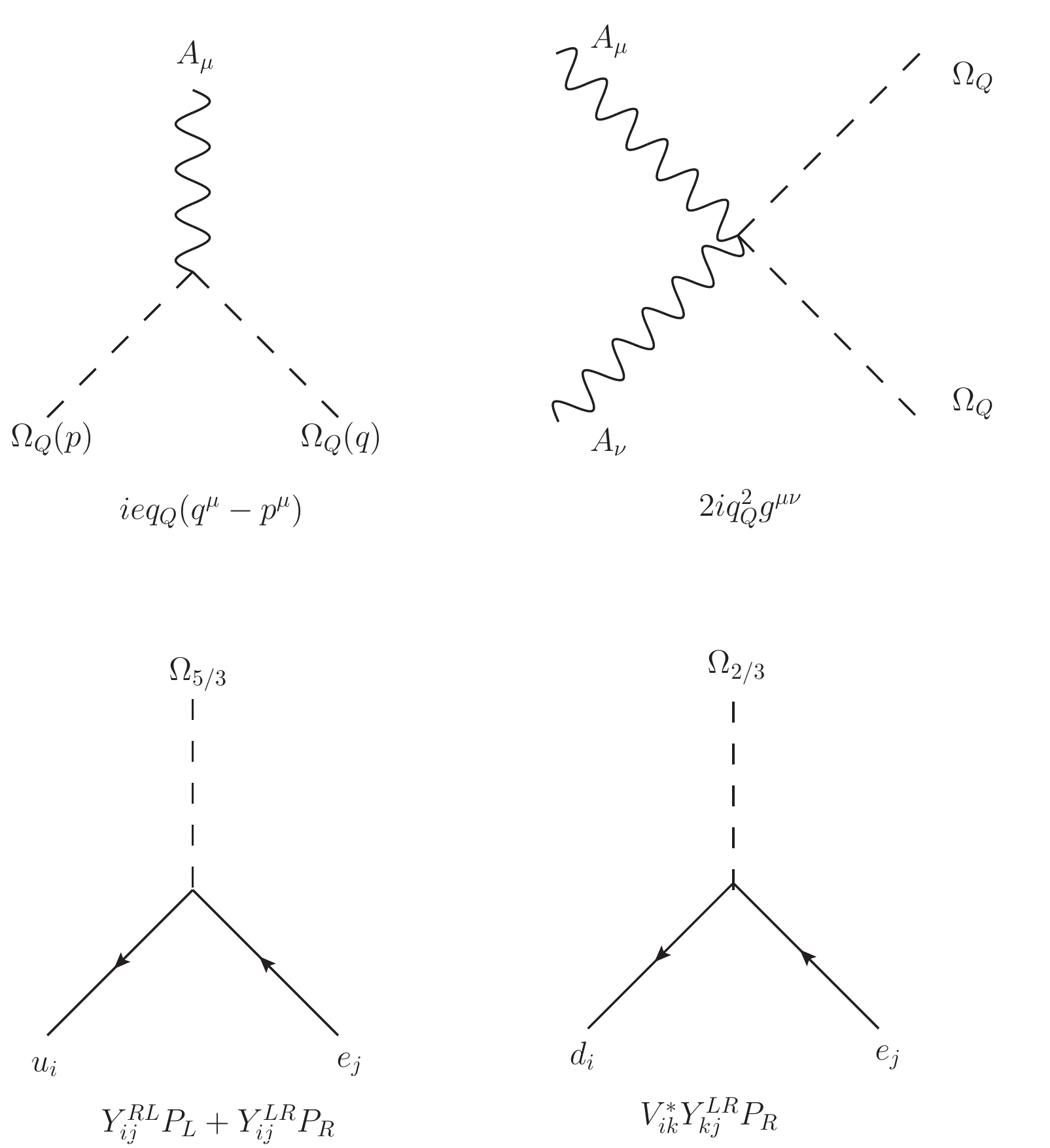}
  \caption{Feynman rules for the $R_2$ scalar LQ interactions to the fermions and photon necessary for the calculation of the $f_i\to f_j\gamma\gamma$ decay.  All the four-momenta are incoming. The usual SM Feynman rules and also that of the propagator of a scalar particle are not included.  \label{FeynmanRules}}
\end{figure}

Apart from the usual SM Feynman rules, the remaining ones necessary for our calculation can be obtained from the above Lagrangians and  are presented in  Fig. \ref{FeynmanRules}. A complete set of Feynman rules for all the $SU(3)_c\times SU(2)_L\times U(1)_Y$ gauge invariant scalar LQ representations are presented in Ref. \cite{Crivellin:2021ejk}.

Below we present the calculations of the three-body decay $t\to c\gamma\gamma$.

\section{One-loop scalar LQ contribution to the $t\to c\gamma\gamma$ decay}
\label{calculation}

 For the sake of completeness we have obtained general results for the contribution of a scalar LQ with charge $Q_S$ to the decay $f_i\to f_j\gamma\gamma$, where $f_{i}$($f_j$) can be a lepton or quark.  These expressions are useful to calculate both the FCNC top quark decay $t\to c\gamma\gamma$ and the LFV decay $\ell_i\to \ell_j\gamma\gamma$ as well.
 It is worth noting that  our results  are also valid for the contribution of other scalar LQs, such as the  weak $SU(2)$ singlet $\chi_{1/3}$.  Although the interaction of such a scalar LQ to a fermion pair involves  Majorana-type Feynman rules that require special treatment, unlike the ones corresponding to the  LQ $\Omega_{5/3}$ and $\Omega_{2/3}$ interactions, it can be shown that after some algebra  the  $\chi_{1/3}$ contributions turn out to be identical to those of $\Omega_{5/3}$ and can be obtained from the latter after replacing  the electric charge $Q_{5/3}\to Q_{1/3}$  and the respective coupling constants to fermion pairs. A similar situation arises in the calculation of the LQ contribution to the two-body decay $f_i\to f_j\gamma$, where the results for $\chi_{1/3}$ can be  obtained from the contribution of $\Omega_{5/3}$ once the corresponding electric charge and coupling constants are replaced.

\begin{figure}[h!]
  \centering
  \includegraphics[width=13cm]{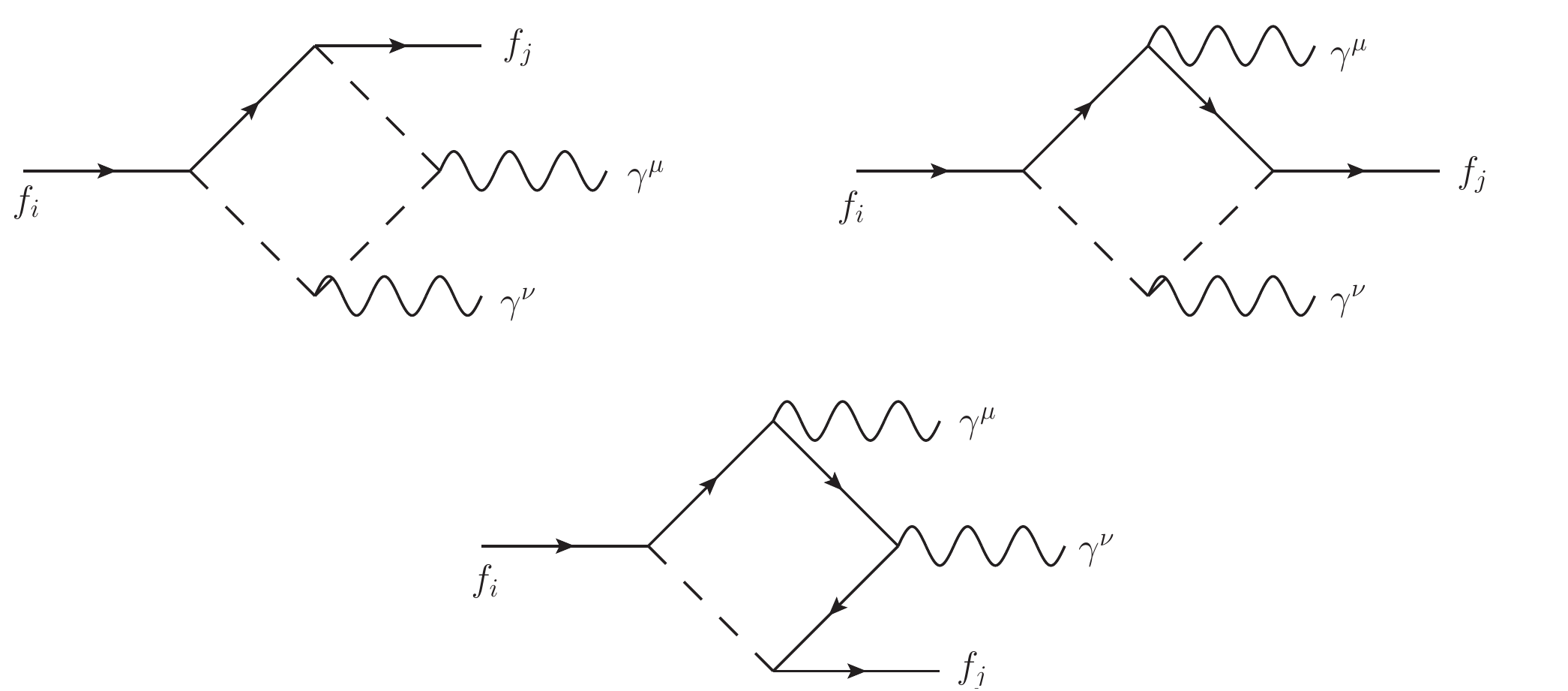}
  \caption{Box diagrams that contribute to the decay $f_i\to f_j\gamma\gamma$ in the LQ model. There are three additional diagrams that are obtained by exchanging the photons.}\label{BoxDiagrams}.
\end{figure}

\begin{figure}[h!]
  \centering
  \includegraphics[width=10cm]{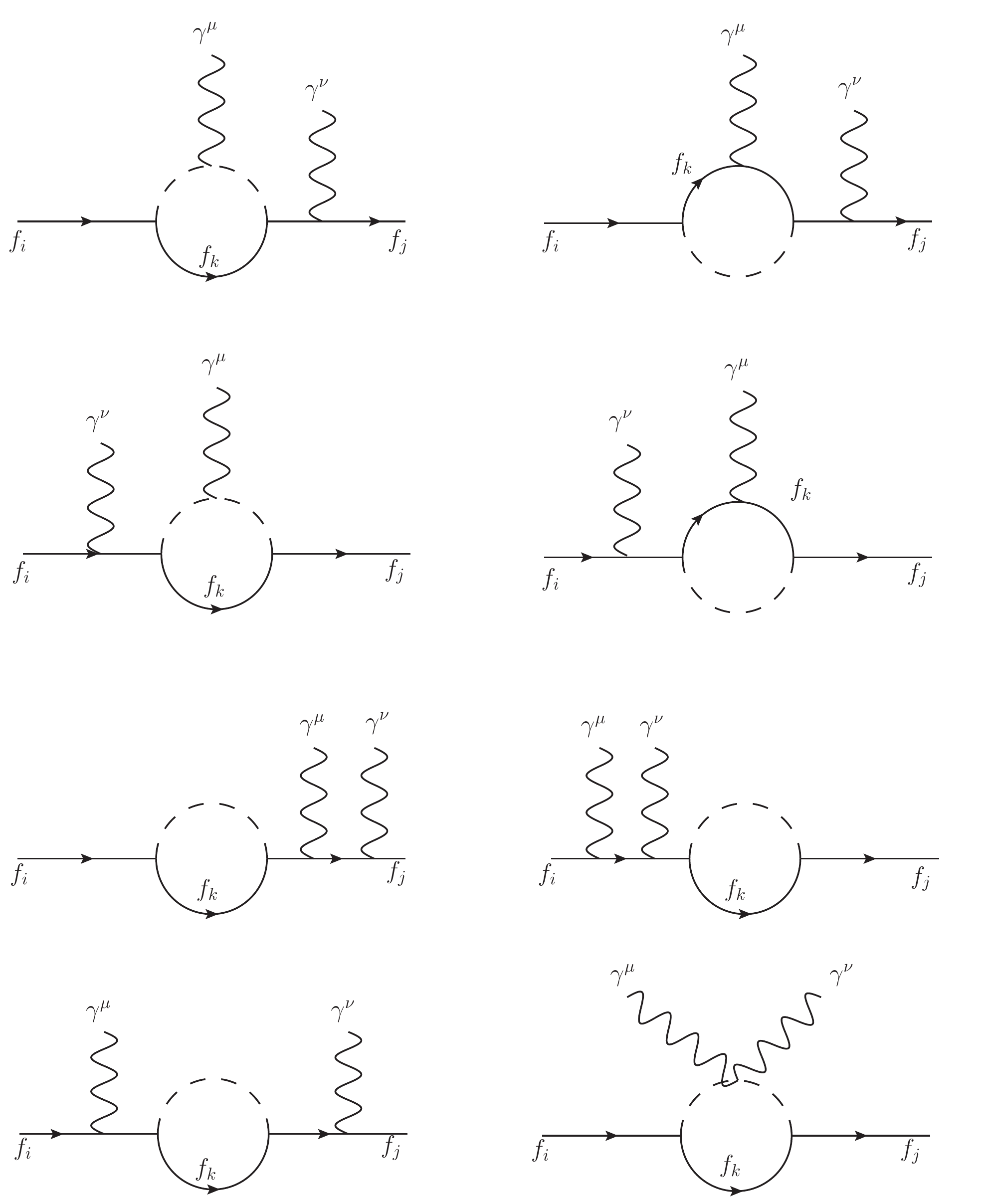}
  \caption{Bubble and triangle Feynman diagrams for the decay $f_i\to f_j\gamma\gamma$ in the LQ model, where $f_i$ and $f_j$ are quarks (charged leptons) and $f_i$ is a lepton (quark). The crossed diagrams that are obtained by exchanging the photons are not shown.}\label{FeynDiagrams1}.
\end{figure}

For our calculation we use the following convention for the particle four-momenta:

\begin{equation}
f_i(p)\to f_j(p')\gamma_\mu(p_1) \gamma_\nu(p_2).
\end{equation}
with $\mu$ and $\nu$ the Lorentz indices of the photon four-momenta $p_1$ and $p_2$. Hence, the mass-shell conditions are given by $p_1^2=p_2^2=0$, $p^2=m_i^2$ and $p'^2=m_j^2$. However  we will use the  $m_j\to 0$ limit as a good approximation since in  all $f_i\to  f_j \gamma\gamma$  decays of phenomenological interest the mass of the outgoing fermion is always negligible as compared to the mass of the ingoing fermion. Also, due to the transversality conditions of the photon fields, i.e., $p_1^\mu \epsilon_\mu(p_1)=p_2^\nu \epsilon_\nu(p_2)=0$, any terms proportional to $p_1^\mu$ and $p_2^\nu$ can be dropped  from the invariant amplitude before contracting with the respective photon polarization vectors. By the same reason,  the replacement $p^\nu\to p_1^\nu+p^{\prime \nu}$ can also be done throughout the calculation.
At the one-loop level, the contribution from  $\Omega_{5/3}$ to the rare three body decay $f_i\to f_j\gamma\gamma$ arises from the box diagrams shown in Fig. \ref{BoxDiagrams} as well as the bubble and triangle diagrams of Fig. \ref{FeynDiagrams1}.

After writing out the invariant amplitude for each Feynman diagram,  the loop integrals were worked out with the Passarino-Veltman  decomposition method
\cite{Passarino:1978jh}. This task was performed with the aid of the Mathematica package FeynCalc \cite{Mertig:1990an,Shtabovenko:2020gxv} and a cross-check was done via  Package-X \cite{Patel:2015tea}. We verified that the amplitude is free of ultraviolet divergences and  obeys both Bose symmetry and  gauge invariance under the $U(1)_\text{em}$ group. It is worth mentioning that ultraviolet divergences cancel out  separately in the  amplitude of each set of Feynman diagrams of Figs. \ref{BoxDiagrams} and \ref{FeynDiagrams1}, whereas  gauge invariance is only achieved after adding up all of the amplitudes.

In the $m_j\to 0$ limit, the invariant amplitude for the decay $f_i\to f_j\gamma\gamma$ can be conveniently written in the following way:

\begin{equation}
\label{invamplitude}
\mathcal M= \epsilon^*_\mu(p_1) \epsilon^*_\nu(p_2)T^{\alpha\mu}(p_1)T^{\beta\nu}(p_2)\bar{f}_j(\mathcal{M}_{L\alpha\beta}P_L+\mathcal{M}_{R\alpha\beta}P_R)f_i,
\end{equation}
where the tensor $T^{\alpha\beta}(p_i)$ is given by
\begin{align}
T^{\alpha\beta}(p_i)&=\frac{1}{m_i^2}\left((p_i\cdot p')g^{\alpha\beta}-p_i^\alpha p'^\beta\right).
\end{align}
which clearly obeys
\begin{align}
T^{\alpha\mu}(p_1)p_{1\mu}&=T^{\alpha\nu}(p_2)p_{2\nu}=0,
\end{align}
and thus  electromagnetic gauge invariance  is manifest.
As far as $\mathcal M_{L\alpha\beta}$ and  $\mathcal M_{R\alpha\beta}$  are concerned, they   are given in terms of six independent form factors $F_n(\hat{s},\hat{t})$
\begin{align}
\label{ML}
\mathcal M_{L\alpha\beta}&=\frac{\alpha N_c}{4\pi m_i}\Bigg[F_1(\hat{s},\hat{t})\, \gamma_\alpha  \gamma_\beta+\frac{1}{m_i}F_2(\hat{s},\hat{t})\, \gamma_\alpha p_{1\beta}+\frac{1}{m_i^2}F_3(\hat{s},\hat{t})\, p_{2\alpha} p_{1\beta}\nonumber\\&+\frac{1}{m_i}\left(F_4(\hat{s},\hat{t})\, \gamma_\alpha \gamma_{\beta}+\frac{1}{m_i}F_5(\hat{s},\hat{t})\, \gamma_\alpha  p_{1\beta}
+\frac{1}{m_i^2}F_6(\hat{s},\hat{t})\, p_{2\alpha} p_{1\beta}\right)(\slashed{p_1}-\slashed{p_2})\Bigg]+\left(\begin{array}{c}p_{1\mu}\leftrightarrow p_{2\nu}\\\hat{s}\leftrightarrow\hat{t}\end{array}\right),
\end{align}
where we introduced the Mandelstam-like scaled variables \begin{align*}
\hat{s}&=\frac{1}{m_i^2}(p-p_1)^2,\\
\hat{t}&=\frac{1}{m_i^2}(p-p_2)^2.
\end{align*}
A similar expression to Eq. \eqref{ML} holds for $\mathcal M_{R\alpha\beta}$ but with $F_n$ replaced by $\widetilde{F}_n$ ($n=1,\ldots 6$), which is obtained from $F_n$ as follows
\begin{equation}
\widetilde{F}_n=F_n\left(\begin{array}{c}Y^{RL}_{i k}\leftrightarrow Y^{LR}_{i k}\\Y^{RL}_{j k}\leftrightarrow Y^{LR}_{j k}\end{array}\right).
\end{equation}
From the above expressions, it is easy to show that Bose symmetry is obeyed.

The contributions of our LQ model to the $F_n(\hat{s},\hat{t})$ form factors are presented in Appendix \ref{FormFactors} in terms of Passarino-Veltman integral coefficients.
The decay width is given by
\begin{equation}
\label{decaywidth}
\Gamma(f_i\to f_j\gamma\gamma )=\frac{m_i}{256\pi^3}\int_0^1 d\hat{s}\int_0^{1-\hat{s}}d\hat{t} \left|\overline{\mathcal{M}}\right|^2,
\end{equation}
where the average square amplitude is presented for completeness in Appendix \ref{squareampli}.

From our general expressions for the decay $f_i\to f_j\gamma\gamma$, we can easily obtain the invariant amplitude and decay width of the $t\to c\gamma\gamma$ process, in which case the internal fermion $f_k$ in the Feynman diagrams of Figs. \ref{BoxDiagrams} and \ref{FeynDiagrams1} is a charged lepton.  We thus set the mass of the decaying and internal fermions as $m_i\to m_t$ and  $m_k=m_\ell$ with $\ell=\mu,\tau$, whereas for the final fermion we use $m_c\simeq 0$. The Mathematica code necessary for all the calculations presented in this work is available for the interested reader in \url{https://github.com/gitavaresve/TopQuarkDecay/tree/main}.

\section{LQ parameter space constraints}
\label{paramcons}
We now discuss the constraints from experimental data  on the parameter space of the model and examine two scenarios: one  with a single LQ doublet and another with multiple LQ doublets.
Below we will concentrate on the bounds on the  LQ masses and the couplings constants $Y^{RL,LR}_{i\ell_j}$, which are required to obtain an estimate for the decay $t\to c\gamma\gamma$.

\subsection{Bounds on LQ mass}
\label{LQmasscons}
 The most up-to-date constraints on the masses of various kinds of vector and scalar LQs have been obtained from the data of direct searches for LQs at the LHC by the CMS and ATLAS collaborations.
Scalar LQs have been searched for through singly production $pp\to S \bar\ell\to q\ell\bar\ell$ or pair production $pp\to S^\dagger S\to q\bar{q}\ell\bar\ell, q\bar q \nu\bar \nu$, thereby yielding bounds on the LQ masses, which depend on the LQ decay channels and the size of their couplings to fermions. Most of these bounds rely on the assumption that LQs can only couple to one generation of fermions and have a dominant decay channel, though very recently a more general scenario where LQs can couple simultaneously to fermions of distinct generations was analysed \cite{ATLAS:2020dsk,ATLAS:2020xov,CMS:2018oaj}. Since a scalar LQ doublet with  non-degenerate mass components can give dangerous contributions to the oblique parameters, we consider that both components of the scalar LQ doublet $R_2$ are mass degenerate, which indeed is true at the lowest order in $\upsilon$. We thus need to consider  the current bounds on the masses of LQs of electric charges $2/3$ and $5/3$.

For the mass of a charge $5/3$ LQ, the most stringent bound was obtained by the CMS collaboration  \cite{CMS:2018svy} by using data collected in 2016 at $\sqrt{s}=13$  TeV and assuming that the main LQ decay mode  is that into a top quark and a tau lepton. Such an analysis has excluded a charge $5/3e$ scalar LQ with a mass below 900 GeV. As for the bounds on other types of scalar LQs, they are more stringent, slightly above 1 TeV. For instance, for the mass of a third-generation charge $2/3e$ scalar LQ decaying into $b\tau/t\nu_\tau$, the ATLAS collaboration \cite{ATLAS:2021jyv} has set a lower  bound   of about 1.2 TeV. As already mentioned, such a charge $2/3$ scalar LQ can be identified with the second  component of the $R_2$ doublet and thus the mass constraint would apply to both $\Omega_{2/3}$ and $\Omega_{5/3}$ if they are considered mass degenerate.

There are also theoretical analyses where the parameter space of scalar LQs have been constrained via the LHC data \cite{Schmaltz:2018nls}. If one considers models where LQs provide an explanation for the LFUV anomalies in $B$ meson decays,   LQ couplings slightly larger than $O(1)$ are required. In this scenario, more stringent constraints on the LQ masses arise from LQ pair production \cite{Angelescu:2018tyl,Angelescu:2021lln}, ranging from 1 to 2 TeVs. In our analysis below we consider LQ masses above 1 TeV and impose constraints on the LQ couplings to fermions such that no dangerous  LQ-mediated contributions to observable quantities are  induced.

\subsection{Bounds on LQ couplings}
\label{LQcoupcons}
In order to discuss the constraints on the LQ couplings we will consider the following two scenarios, which can allow us to asses the order of magnitude of the   $t\to c\gamma\gamma$ branching ratio.

{\bf Scenario I)}There is only one LQ doublet $R_2$ that has both left- and right-handed couplings to fermions of the second and third generations only, namely, $Y^{RL,LR}_{i\ell}$, where $\ell=\mu,\tau$ and $i=2,3$ stands for the quark generation. In this scenario there are LFV transitions between the muon and the tau  lepton and there  is  indeed an explanation for the muon $g-2$ anomaly,  though an explanation for the apparent LFUV anomalies in $B$ meson decays is not  favored.  This model was considered in our previous work on the two-body top quark decays $t\to cX$ ($X=\gamma,g,H,Z$) \cite{Bolanos:2019dso}  and constraints on the parameter space were obtained from the muon $g-2$ anomaly and the LFV decay $\tau\to \mu\gamma$, together with extra constraints  to avoid large contributions to other observable quantities.

{\bf Scenario II)}There are multiple LQ doublets  $R_2$. As an example of this realization we will consider the model recently proposed in \cite{Crivellin:2022mff}, where the SM is augmented with one LQ doublet $R_2^\ell$ $(\ell=e,\mu,\tau)$ for each lepton generation. Although each LQ doublet can only couple to the leptons of one generation, thereby forbidding LFV, all of them can couple  to the quarks of the second and third generations. This scenario provides an explanation for  the $a_\mu$ anomaly and  the  LFUV anomalies in $B$ meson decays, which requires relatively large LQ couplings, though as mentioned above recent data seems to exclude the $R_{K}$ and $R_{K^*}$ anomalies.  An analysis on the constraints on the parameter space of this model was presented in Ref. \cite{Crivellin:2022mff}.

Below we analyse the constraints on the LQ couplings in the above scenarios and focus on the allowed parameter space region most promising  for the $t\to c\gamma\gamma$  branching ratio.

\subsubsection{Scenario I}
In our analysis we will follow our previous work \cite{Bolanos:2019dso}, where we assumed that the  $\Omega_{5/3}$ scalar LQ is responsible for the muon $g-2$ anomaly and considered the bounds on the $Y^{LR,RL}_{i\mu }$ and $Y^{LR,RL}_{i\tau }$ couplings obtained from the LFV decay $\tau\to \mu\gamma$.
The analytical expressions for the contribution of a scalar LQ  to the muon anomalous magnetic dipole moment (AMDM) and the LFV decay $\ell_i\to \ell_j\gamma$ were obtained  long ago and  were also reproduced in Ref. \cite{Bolanos:2013tda,Bolanos:2019dso} in terms of Feynman parameter integrals and Passarino-Veltman scalar functions. For the sake of completeness we present such results in Appendix \ref{LFVformulas}. Note that   $a_\mu$ contains a chirality flipping term proportional to  $m_{q_i}\times {\rm Re}\left(Y_{i\mu}^{RL} Y_{i\mu}^{LR*}\right)$, which gives the dominant contribution for a heavy internal quark. Since such contribution requires that the scalar LQ has both left- and right-handed couplings to the fermions, it is absent for chiral LQs. Thus the contribution to $a_\mu$ from the chiral LQ $\Omega_{2/3}$  via an internal $b$ quark is expected to be much smaller than the contribution from the non-chiral LQ $\Omega_{5/3}$.

As far as the experimental constraints are concerned, for the muon AMDM $a_\mu$ we consider the current  average of its experimental measurements \cite{Muong-2:2006rrc,Muong-2:2021ojo}, whereas for the SM theoretical prediction we consider the   estimate obtained by the muon $g-2$ theory initiative \cite{Aoyama:2020ynm}. This yields the following $4.2\sigma$ discrepancy between theory and experiment:
\begin{equation}
\Delta a_\mu=251 (59)\times 10^{-11}.
\end{equation}
For the decay $\tau\to \mu\gamma$, the expected future experimental sensitivity \cite{Belle-II:2018jsg}  puts strong constraints on LFV processes between the second and third generations
\begin{equation}
{\rm Br}(\tau\to \mu\gamma)\le 1.0\times 10^{-9},
\end{equation}
but we will consider the current experimental constraint \cite{BaBar:2009hkt}
\begin{equation}
{\rm Br}(\tau\to \mu\gamma)\le 4.4\times 10^{-8}.
\end{equation}

Note that  $\Omega_{5/3}$ and $\Omega_{2/3}$ can also give dangerous contributions to several observable quantities  through their couplings to the $\bar b\tau$ and $t \nu_\tau$ pairs both at the tree and one-loop level.  Therefore one must verify that the values of the LQ couplings consistent with the muon $g-2$ discrepancy and the constraint on the $\tau\to\mu\gamma$ decay do not introduce tension between the theory predictions and  experimental data. For instance at the LHC, double (single) tau lepton production $\tau^- \tau^+$ ($\tau \nu$)   can arise at the tree level via single LQ production $pp\to S\ell^+\to q\ell^-\ell^+$ ($pp\to S\ell^+\to q\nu\ell^+$), with $q$ being identified with a single jet. On the other hand,  among those observable quantities sensitive to large LQ couplings at the one-loop level there are the decays $Z\to \tau^-\tau^+$ and $Z\to \nu\nu$. We will thus impose the additional constraints $|Y^{LR,RL}_{i\mu}|<1$, which ensures that   no tension will arise  between the theory predictions and the experimental data \cite{Angelescu:2021lln,Crivellin:2022mff}. This is a more stringent constraint  than  $|Y^{LR,RL}_{i\mu}|<4\pi$, which is usually imposed  to avoid the breakdown of perturbation theory.

\begin{figure}[!htb]
\includegraphics[width=18cm]{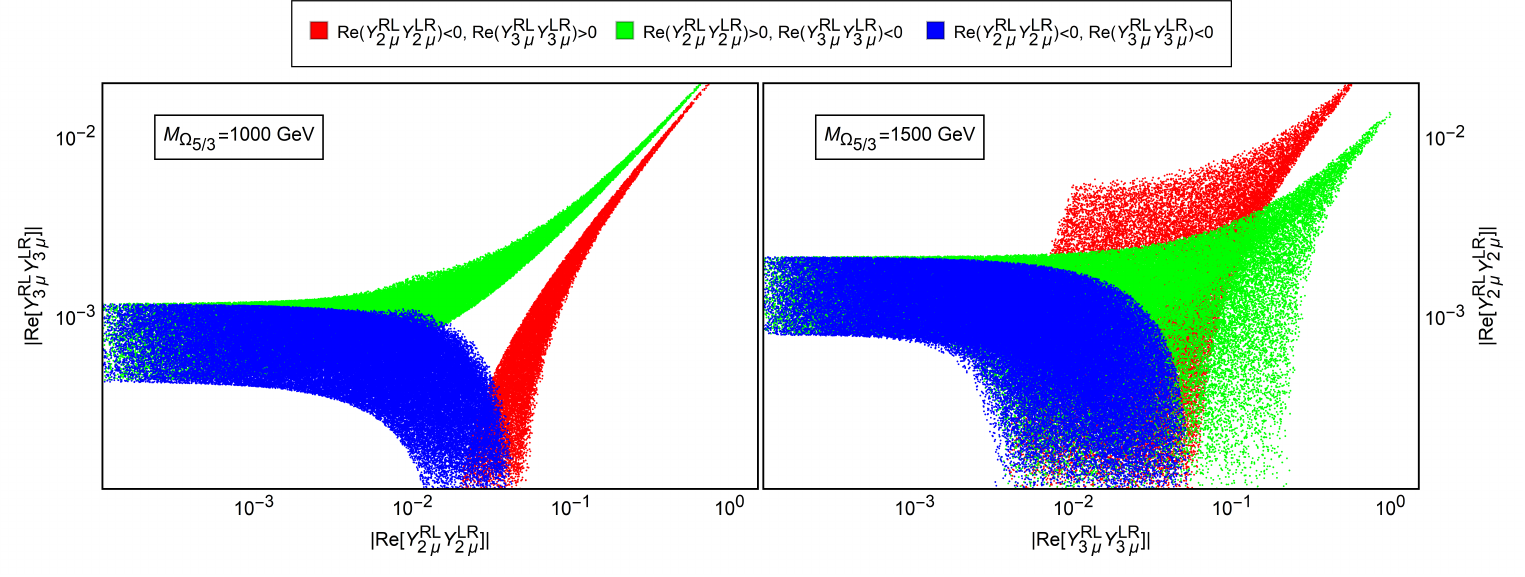}
\caption{Allowed area on the  ${\rm Re}\left(Y_{2\mu }^{RL} Y_{2\mu }^{LR}\right)$ vs ${\rm Re}\left(Y_{3\mu }^{RL} Y_{3\mu}^{LR}\right)$ plane consistent with the muon $g-2$ anomaly at 95\% C.L. for  a charge $5/3$ scalar LQ with mass of 1 and 1.5 TeVs in three scenarios of the signs of the ${\rm Re}\left(Y_{i\mu}^{RL} Y_{i\mu}^{LR}\right)$ products, which determine the signs of the partial contributions. As explained in the text, we also impose the extra constraints $|Y^{LR,RL}_{i\mu}|<1$ to be consistent with other constraints from experimental data.    \label{amubounds}}
\end{figure}

Rather than making any a priori assumption about the LQ coupling constants to leptons and quarks, we consider non-vanishing left- and right-handed couplings  to the leptons and quarks of both the second and third generations: $Y^{LR,RL}_{i\mu}$, where $i$ stands for the generation quark. Without losing generality we consider purely real LQ couplings and randomly scan  for a few thousands of 4-tuples $\{Y_{2\mu }^{RL},Y_{2\mu }^{LR},Y_{3\mu }^{RL},Y_{3\mu }^{LR}\}$ consistent with the discrepancy of the muon $g-2$ anomaly at 95\% C.L. The allowed region is shown in Fig. \ref{amubounds} on the ${\rm Re}\left(Y_{2\mu }^{RL} Y_{2\mu }^{LR}\right)$ vs ${\rm Re}\left(Y_{3\mu }^{RL} Y_{3\mu}^{LR}\right)$ plane for a charge $5/3$ scalar LQ  with a mass of 1 TeV and $1.5$ TeV.

In our analysis we have considered three scenarios for the relative signs of the products of the left- and right-handed couplings  ${\rm Re}\left(Y_{i\mu}^{RL} Y_{i\mu}^{LR}\right)$ as they determine  the sign of  the contribution to $a_\mu$ from the quark of generation $i$. We observe that the largest  values for the coupling products are allowed when  they have opposite signs (red and green points) as large partial contributions can cancel each other out to give the required negative total contribution. On the other hand, in the scenario when both couplings  are negative (blue points), the partial contributions add up, thereby imposing a tighter constraint of the LQ coupling products.

\begin{figure}[!htb]
\includegraphics[width=18cm]{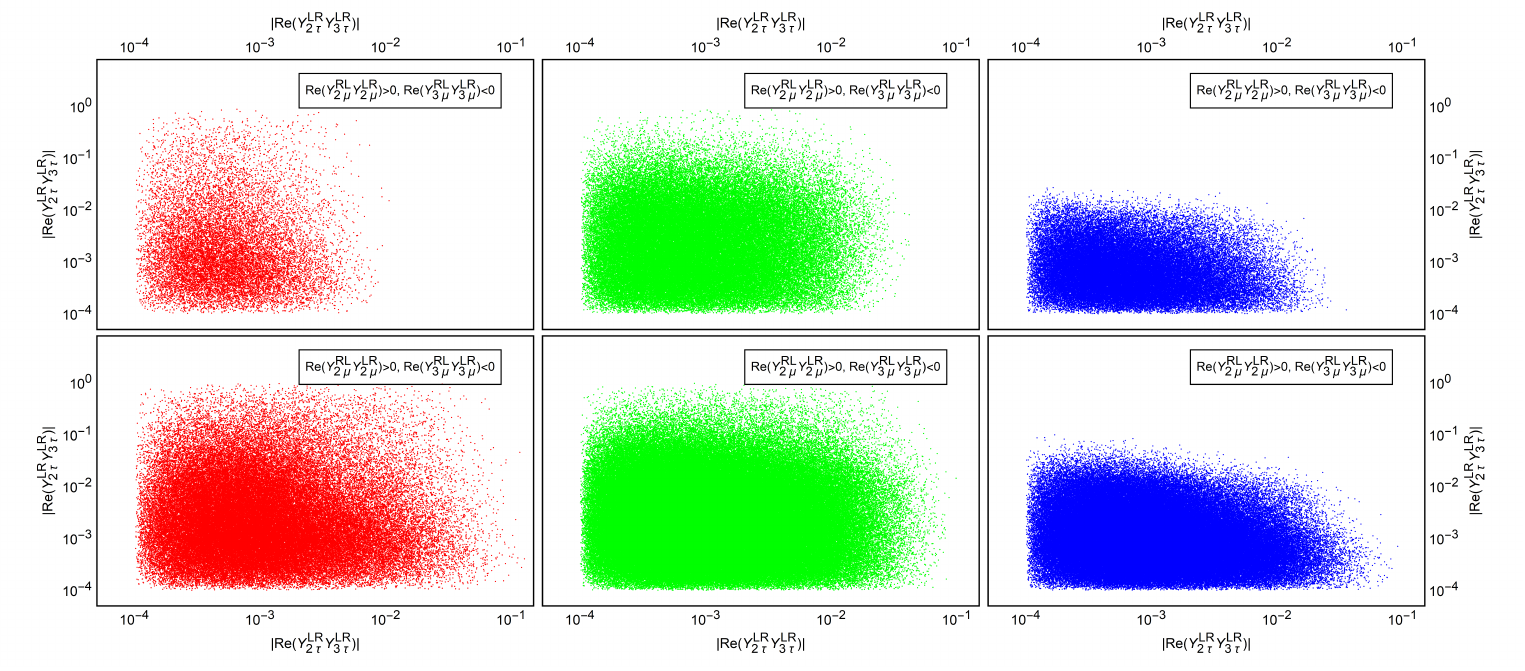}
\caption{Allowed area on the ${\rm Re}\left(Y_{2\tau }^{RL} Y_{3\tau}^{RL}\right)$ vs ${\rm Re}\left(Y_{2\tau }^{LR} Y_{3\tau }^{LR}\right)$  plane consistent with the muon $g-2$ anomaly at 95\% C.L. and the LFV decay $\tau\to\mu\gamma$ for  a scalar LQ $\Omega_{5/3}$  with mass of 1 TeV (top plots) and 1.5 TeV (bottom plots) in the three scenarios of the signs of the products  ${\rm Re}\left(Y_{i\mu}^{RL} Y_{i\mu}^{LR}\right)$ considered in Fig \ref{amubounds}. The constraint  $|Y_{i \ell }^{RL,LR}|\le 1$ is also imposed to be consistent with other experimental constraints. \label{lfvbounds}}
\end{figure}

\begin{figure}[!ht]
\includegraphics[width=18cm]{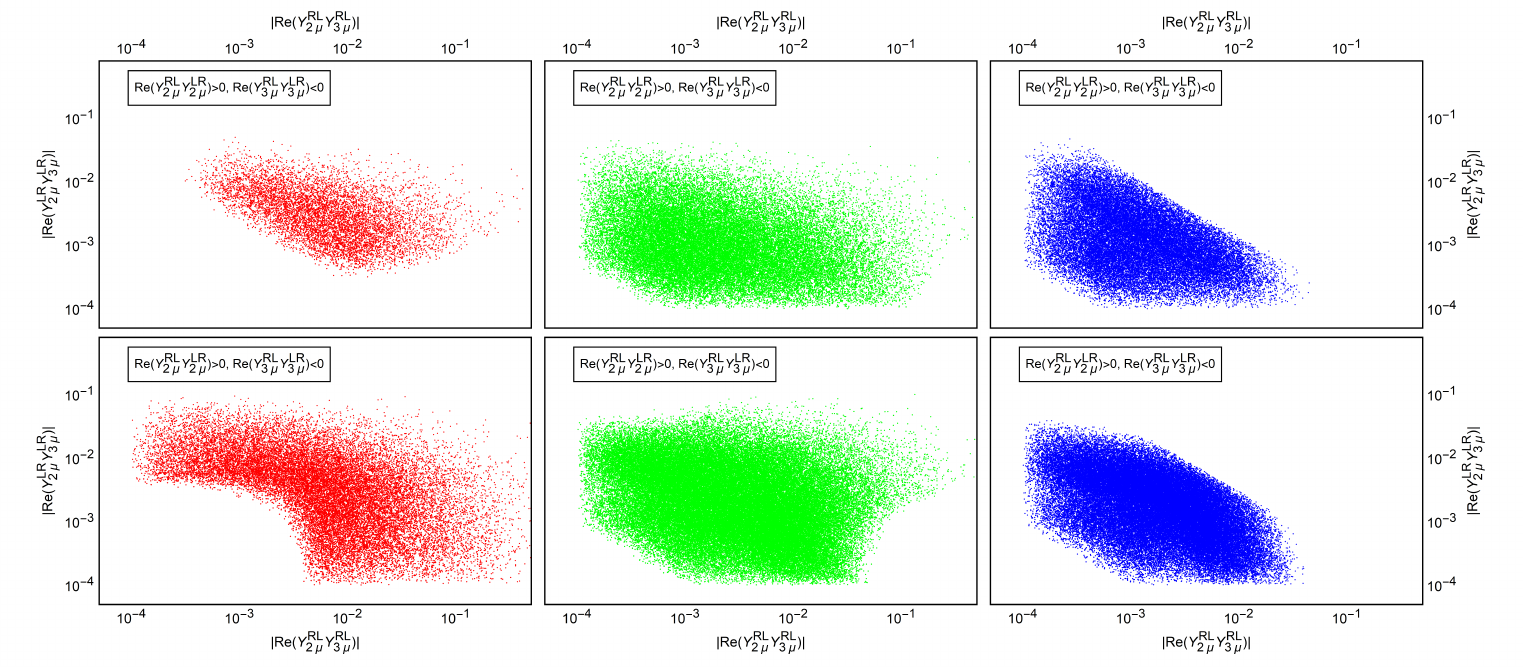}
\caption{The same as in Fig. \ref{lfvbounds} but for the allowed area  on the ${\rm Re}\left(Y_{2\mu }^{RL} Y_{3\mu}^{RL}\right)$ vs ${\rm Re}\left(Y_{2\mu }^{LR} Y_{3\mu }^{LR}\right)$  plane.  \label{lfvbounds1}}
\end{figure}

Unless an additional flavor symmetry is introduced,   $\Omega_{5/3}$  can also couple to the $\tau$ lepton and induce LFV processes.
Thus the decay $\tau\to \mu\gamma$ imposes an extra constraint on the $Y_{i\mu}^{RL,LR}$ and  $Y_{i\tau}^{RL,LR}$ couplings, which must be combined with that arising from the muon $g-2$ anomaly. Again we do not impose an a priori condition for the coupling constants and  randomly scan for set of points  $\{Y_{2\ell}^{RL},Y_{2\ell}^{LR},Y_{3\ell}^{RL},Y_{3\ell}^{LR}\}$ ($\ell=\mu,\tau$) consistent with the discrepancy of the muon $g-2$ anomaly  and the experimental upper bound on the $\tau\to \mu\gamma$ decay, along with the extra upper bounds $|Y_{i\ell}^{RL,LR}|\le 1$. It turns out that the $t\to c\gamma\gamma$ invariant amplitude of Eq. \eqref{invamplitude} is given in terms of form factors of the form (see Appendix \ref{FormFactors})

\begin{align}
F_n
      &=\sum_{\ell_k=\mu,\tau}\left(f_n^{LL} Y^{RL}_{i\ell_k} Y^{RL}_{j\ell_k}+f_n^{RR} Y^{LR}_{i\ell_k} Y^{LR}_{j\ell_k}+f_n^{RL} Y^{LR}_{i\ell_k} Y^{RL}_{j\ell_k}
      +f_n^{LR}Y^{RL}_{i\ell_k} Y^{LR}_{j\ell_k}\right).
      \label{Fn}
\end{align}
From our numerical analysis of the allowed values for the LQ coupling constants, we infer that from all the products of coupling constants of Eq. \eqref{Fn}, the ones that can reach the largest allowed values are $Y^{RL}_{3\mu} Y^{RL}_{2\mu}$ and $Y^{LR}_{3\tau} Y^{LR}_{2\tau}$, whereas the remaining products are much more constrained. Therefore the $t\to c\gamma\gamma$ decay width will depend mainly on the values of this pair of coupling products and  would reach its maximal value when one of them reaches its largest allowed values.  We thus show  the allowed region on the ${\rm Re}\left(Y_{2\tau }^{RL} Y_{3\tau}^{RL}\right)$ vs ${\rm Re}\left(Y_{2\tau }^{LR} Y_{3\tau }^{LR}\right)$  plane in  Fig. \ref{lfvbounds} and  on the  ${\rm Re}\left(Y_{2\mu }^{RL} Y_{3\mu}^{RL}\right)$ vs ${\rm Re}\left(Y_{2\mu }^{LR} Y_{3\mu }^{LR}\right)$  plane in Fig. \ref{lfvbounds1} for  a scalar LQ $\Omega_{5/3}$  with mass of 1 TeV (top plots) and 1.5 TeV (bottom plots) in the three scenarios of the signs of the products  ${\rm Re}\left(Y_{i\mu}^{RL} Y_{i\mu}^{LR}\right)$ of Fig. \ref{amubounds}.

We can conclude that the largest allowed values correspond to  $Y_{2\tau }^{LR} Y_{3\tau }^{LR}$ products, which  are obtained when both the $c$ and $t$ contributions to $a_\mu$ have opposite signs (red and green points). The products of LQ couplings to the muon  are more restricted, which is due to the condition imposed by the muon $g-2$ anomaly.
 For illustration purpose, in Table \ref{allowpoints} we show a few  sets of  values in which the product $Y^{LR}_{2\tau} Y^{LR}_{3\tau}$ reaches its largest allowed values, which are slightly below  the unity. These sets of points  would yield the maximal values of the $t\to c\gamma\gamma$ branching ratio in scenario I.

Also, as expected we observe   that the constraints on the LQ couplings are relaxed when the LQ mass increases and the allowed area enlarges slightly when the LQ increases from 1 to 1.5 TeVs. However, although the LQ couplings could be less restricted, a possible enhancement of the  $t\to c\gamma\gamma$  branching ratio may be suppressed by the larger value of the LQ mass.

\subsubsection{Scenario II}
Motivated by the apparent anomalies in $B$ meson decays, quite recently the authors of Ref. \cite{Crivellin:2022mff} introduced a model with one LQ doublet $R_2^\ell$ ($\ell=\mu,\tau$) for each lepton generation.  In order to avoid dangerous LFV effects induced by  large LQ couplings, required by the LFUV $B$ meson anomalies, an additional flavor symmetry was imposed so that each scalar LQ doublet  can only couple to one lepton generation, though they still have couplings to quarks of the second and third generations. Therefore, although no LQ-mediated LFV processes arise,  FCNC top quark decays could still be allowed at the one-loop level.

A comprehensive analysis of the bounds on LQ $R_2^\tau$ couplings obtained from several observable quantities was presented in Ref. \cite{Crivellin:2022mff}.
It was shown that one scalar LQ doublet  $R_2^\tau$ with a mass of 1.7 TeVs can  explain the $R_D$ and $R_{D^*}$ anomalies, provided that the values of the coupling constants $Y_{3\tau}^{LR}$ and $Y_{2\tau}^{RL*}$ are slightly larger than $O(1)$ and there is a large complex phase in the product of both coupling constants. This is still consistent with the data on  $\tau \nu$ and $\tau^- \tau^+$ production at the LHC, but some tension could arise in the electroweak fit for the data of the decays $Z\to \tau^-\tau^+$ and $Z\to \bar{\nu}\nu$ \cite{Crivellin:2022mff}. Also, the constraint $Y_{2\tau}^{LR} Y_{3\tau}^{LR*}\lesssim 0.25$ is obtained from  $B_s-\bar{B}_s^*$ mixing  \cite{Crivellin:2022mff}.

The $R_2^\tau$ doublet alone gives no explanation for both the muon $g-2$ discrepancy and the apparent $R_{K}$ and $R_{K^*}$ anomalies, for which  a second doublet LQ doublet  $R_2^\mu$ is necessary \cite{Crivellin:2022mff}.
It was shown that the presence of two scalar doublets improves the fit of the $b\to s\ell^-\ell^+$ data and relax the tension in the electroweak data as smaller $R_2^\tau$ couplings are required to explain the $R_D$ and $R_{D^*}$ anomalies for $M_{R_2^\mu}=2$ TeV and $M_{R_2^\tau}=1.7$ TeV.  A summary of the values of the coupling constants consistent with the LFUV anomalies in $B$ decays and the $a_\mu$ anomaly, taken from Ref. \cite{Crivellin:2022mff}, is shown in the last row of Table \ref{allowpoints}. Note that the LQ doublet  $R_2^\tau$ alone could yield an enhanced branching ratio for  FCNC top quark decays due to the large LQ couplings, though  an additional suppression due to the heavier value of the  LQ mass is expected too.

\begin{table}[!ht]
\caption{Sample sets of values of LQ  couplings to fermions consistent with the muon $g-2$ anomaly and
the experimental constraint on the LFV decay $\tau\to\mu\gamma$ for $m_{R_2}=1$ TeV (rows 1 through 3) and $m_{R_2}=1.5$ TeV (rows 4 through 7) in scenario I. The last row shows the allowed values consistent with an explanation for the  $R_{D}$ and $R_{D^*}$ anomalies, the muon $g-2$ discrepancy, and other experimental constraints for $m_{R_2^\mu}=1.7$ TeV and $m_{R_2^\mu}=2$  TeV in scenario II (see \cite{Crivellin:2022mff}), which  also explains the apparent $R_{K}$ and $R_{K^*}$ anomalies, which seem to be excluded by recent data \cite{LHCb:2022qnv}. Results  taken from Ref.   \cite{Crivellin:2022mff}.\label{allowpoints}}
\renewcommand{\arraystretch}{1.4}
\begin{tabular}{cccccccccc}
\hline
\hline
Scenario&LQ mass (TeV)  &$Y_{2\mu}^{RL}$& $Y_{2\mu}^{LR}$&$Y_{3\mu}^{RL}$& $Y_{3\mu}^{LR}$&$Y_{2\tau}^{RL}$& $Y_{2\tau}^{LR}$
&$Y_{3\tau}^{RL}$& $Y_{3\tau}^{LR}$\\
\hline
\hline
I&1.0&-0.167 & 0.246 & 0.011 & 0.018 & 0.059 & 0.92 & 0.015 & 0.88\\
I&1.0& 0.949 & 0.013 & -0.073 & 0.018 & 0.011 & 0.917 & 0.018 & 0.798 \\
I&1.0& 0.193 & 0.019 & -0.014 & 0.044 & 0.024 & 0.978 & 0.012 & 0.888 \\
I&1.5&0.431 & 0.041 & -0.025 & 0.011 & 0.037 & 0.994 & 0.016 & 0.831 \\
I&1.5& 0.424 & 0.012 & -0.02 & 0.021 & 0.023 & 0.957 & 0.041 & 0.917 \\
I&1.5&-0.236 & 0.155 & 0.012 & 0.038 & 0.016 & 0.958 & 0.013 & 0.995 \\
II&$1.7$ $(R^\tau)$, $2.0$ $(R^\mu)$&--&--&2.4-2.6&$-7\times 10^{-3}$&2.4-2.55&0.3-0.35&--&0.9-1.1\\
\hline
\end{tabular}
\end{table}

Following the notation of  Ref. \cite{Angelescu:2018tyl} we present in Table \ref{modelfea} the main features of the scenarios just discussed.  It seems that  scenario II is the most promising for a less suppressed $t\to c\gamma\gamma$  branching ratio as LQ couplings to the $\tau$ lepton  of the order of $O(1)$ are  allowed, however, a larger mass is also necessary to fulfil the constraints from experimental data, which  may result in an additional suppression. Below we present an estimate for the $t\to c\gamma\gamma$  branching ratio in these scenarios.

\begin{table}[!ht]
\caption{Scenarios discussed in the text along with the anomalies addressed and other  predicted new physics effects. Note that the $R_{K^*}$ anomaly seems to be excluded by the recent LHCb measurement \cite{LHCb:2022qnv}.\label{modelfea}}
\renewcommand{\arraystretch}{1.4}
\begin{tabular}{lccccc}
\hline
\hline
Scenario &$a_\mu$&$R_D$, $R_{D^*}$&$R_{K^*}$&LFV&FCNC\\
\hline
\hline
I&\Checkmark&\XSolidBrush&\XSolidBrush&\Checkmark&\Checkmark\\
II with  $R_2^\tau$ alone&\XSolidBrush&\Checkmark&\XSolidBrush&\XSolidBrush&\Checkmark\\
II with both $R_2^\mu$ and $R_2^\tau$&\Checkmark&\Checkmark&\Checkmark&\XSolidBrush&\Checkmark\\
\hline
\end{tabular}
\end{table}

\subsection{Estimate of the $t\to c\gamma\gamma$ branching ratio}
For the numerical evaluation of the invariant amplitude \eqref{invamplitude} and to achieve a best numerical stability in the evaluation of the double integral of Eq. \eqref{decaywidth}, we decomposed the Passarino-Veltman integrals of Eqs. \eqref{f1LL} through \eqref{f6RR}  into scalar functions (the results are too lengthy to  be presented in this work), which then were evaluated through the LoopTools \cite{vanOldenborgh:1990yc,Hahn:1998yk} package. Also, an independent evaluation was performed via the  Collier package \cite{Denner:2016kdg}, which showed a good agreement with the LoopTools evaluation. To obtain de $t\to c\gamma\gamma$ decay width we impose the kinematic cuts $E_\gamma>5$ GeV for both photons.

As far as scenario I is concerned, from the above analysis we can conclude that the contribution to the $t\to c\gamma\gamma$ decay width from the loops with an internal  muon is much smaller than that from the loops with an internal tau lepton, which is due to the small allowed values of the LQ-muon couplings. In fact we can neglect the muon contribution to the  $t\to c\gamma\gamma$ invariant amplitude as it is more than two orders of magnitude below than the one of the tau lepton. Therefore, the largest values of the  $t\to c\gamma\gamma$ decay width for a specific value of the LQ mass are  reached in the region where the  product of couplings $Y^{LR}_{2\tau} Y^{LR}_{3\tau}$  reaches its largest allowed values. In this region the behaviour of the $t\to c\gamma\gamma$ branching ratio as a function of the LQ coupling can be roughly approximated as
\begin{equation}
Br(t\to c\gamma\gamma)\simeq f(m_S,m_c,m_t,m_\tau) |Y^{LR}_{2\tau} Y^{LR}_{3\tau}|^2,
\end{equation}
with $f(m_S,m_c,m_t,m_\tau)$ of the order of $10^{-11}$--$10^{-10}$ at most for $m_S\simeq 1$ TeV.

\begin{figure}[!ht]
\includegraphics[width=10cm]{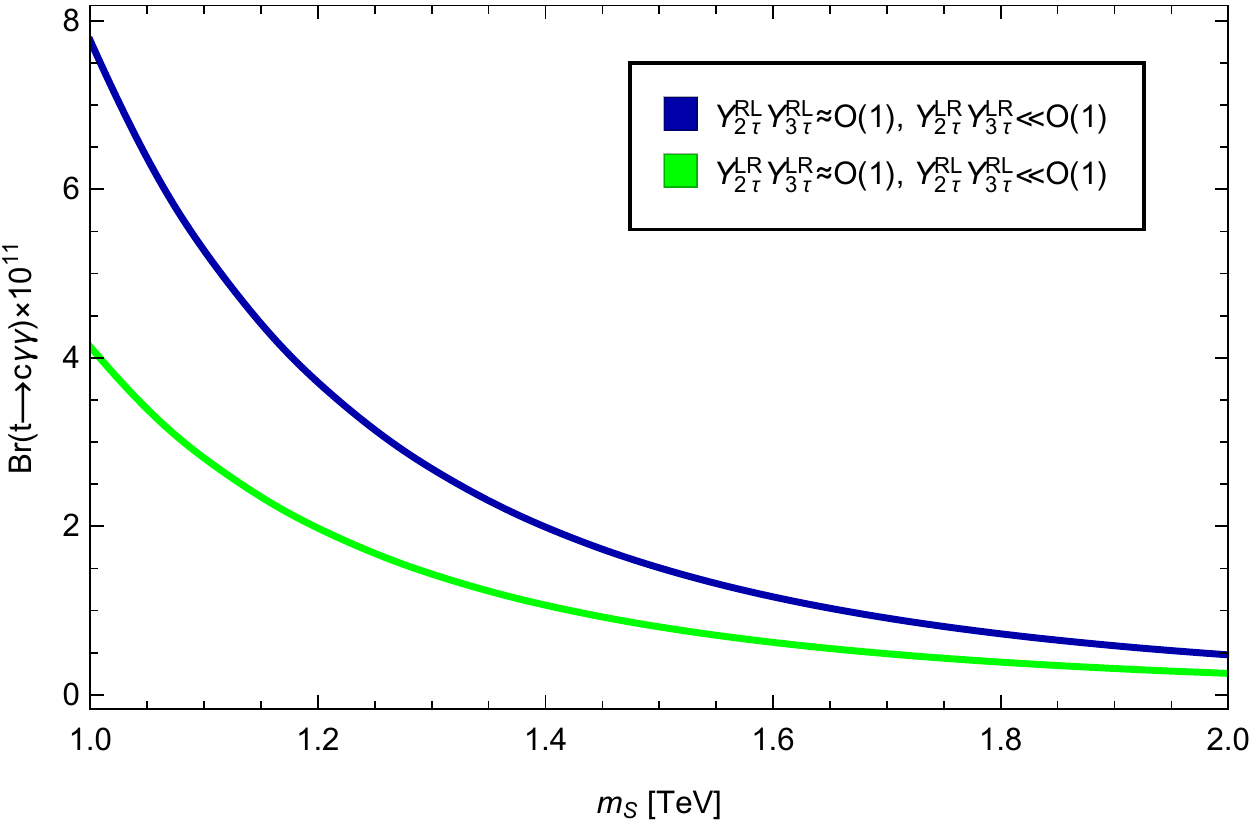}
\caption{LQ contribution to the branching ratio of the $t\to c\gamma\gamma$ decay in scenario I as a function of the LQ mass   in the region of the parameter space where the product $Y^{LR}_{2\tau} Y^{LR}_{3\tau}\sim O(1)$  reaches its largest allowed values, which corresponds to the largest possible value of $Br(t\to c \gamma\gamma)$. We also include the case where the product $Y^{RL}_{2\tau} Y^{RL}_{3\tau}$ dominates over $Y^{LR}_{2\tau} Y^{LR}_{3\tau}$ for $Y^{RL}_{2\tau} Y^{RL}_{3\tau}\simeq O(1)$, though this is not allowed by the constraints from experimental data.  \label{branchingplot}}
\end{figure}

In Fig. \ref{branchingplot} we show the behaviour of the LQ contribution to $Br(t\to c\gamma\gamma)$ in scenario I  as a function of the LQ mass around the region where the $Y^{LR}_{2\tau} Y^{LR}_{3\tau}$ product reaches its largest allowed values and thus  $Br(t\to c\gamma\gamma)$ reaches its largest values. For illustration purpose we also show  the case where the product $Y^{RL}_{2\tau} Y^{RL}_{3\tau}$ would dominate over $Y^{LR}_{2\tau} Y^{LR}_{3\tau}$, though in this case $Br(t\to c\gamma\gamma)$ does not reach its largest values as  $Y^{RL}_{2\tau} Y^{RL}_{3\tau}$ is  allowed to be of the order of $10^{-1}$ at most.

We can conclude that the LQ contribution  to $Br(t\to c\gamma\gamma)$ can be of the order of $10^{-11}$ at most for $Y^{LR}_{2\tau} Y^{LR}_{3\tau}$ of the order of $O(1)$, though it would decrease considerably if the LQ couplings decrease by one order of magnitude.  Also, we note that there is little dependence on the LQ mass in the interval from 1 to 2 TeVs.

As for scenario II, according to the allowed values of the LQ couplings presented in Table \ref{allowpoints} from Ref. \cite{Crivellin:2022mff}, $Br(t\to c\gamma\gamma)$ would be dominated by the $R_2^\tau$ contribution since any products of $R_2^\mu$ couplings that enter into Eq. \eqref{Fn} would be considerably suppressed.  Furthermore, the dominant term arises from the  $Y^{RL}_{2\tau} Y^{LR}_{3\tau}$ product. In this case we obtain for $m_{R_2^\tau}=1.7$ TeV
\begin{equation}
Br(t\to c\gamma\gamma)\simeq 9.11\times 10^{-12}\times| Y^{RL}_{2\tau} Y^{LR}_{3\tau}|^2,
\end{equation}
where we have neglected all other contributions. For $Y^{RL}_{2\tau}\simeq 2$ and $Y^{LR}_{3\tau}\simeq 1$ we obtain again the estimate $Br(t\to c\gamma\gamma)\simeq 10^{-11}$. Therefore, in both scenario I and scenario II, the $t\to c\gamma\gamma$ branching ratio would reach values as high as $10^{-12}$--$10^{-11}$. These are the largest possible values that one can expect for the contribution to the $t\to c \gamma\gamma$ decay from the $R_2$ doublet scalar LQs since experimental constraints severely constraint the LQ coupling constants. Thus, unless an extraordinary cancellation  in the LQ contribution to experimental observable quantities (fine tuning) occurs in a more sophisticated framework with additional LQ multiplets, the LQ contribution to the $t\to c\gamma\gamma$ decay is expected to be beyond the reach of measurement.

In Table \ref{branchings} we present a summary of the one-loop contributions to two- and three-body FCNC top quark decays  from a scalar $SU(2)$ doublet from Ref. \cite{Bolanos:2019dso} and this work. We have considered the largest estimate consistent with the current constraints from experimental data.

\begin{table}[!htb]
\caption{Largest estimated values of the one-loop contributions to two- and three-body FCNC top quark decays  from  models with scalar $SU(2)$ doublets consistent with the current constraints from experimental data \cite{Bolanos:2019dso}.\label{branchings}}
\renewcommand{\arraystretch}{1.4}
\begin{tabular}{ccc}
\hline
\hline
Decay channel&Branching ratio &Branching ratio \\
&$m_{R_2}=1$ TeV&$m_{R_2}=1.5$ TeV\\
\hline
\hline
$t\to c\gamma$&$10^{-9}$&$10^{-9}$\\
$t\to cg$&$10^{-9}$&$10^{-10}$\\
$t\to cZ$&$10^{-8}$&$10^{-9}$\\
$t\to cH$&$10^{-9}$&$10^{-10}$\\
$t\to c\mu^-\mu^+$&$10^{-6}$&$10^{-7}$\\
$t\to c\tau^-\tau^+$&$10^{-7}$&$10^{-8}$\\
$t\to c\gamma\gamma$&$10^{-11}$&$10^{-12}$\\
\hline
\end{tabular}
\end{table}

%\subsection{The $\tau\to \mu\gamma\gamma$ decay}
%As a by-product, from our results we can easily estimate the branching ratio for the $\tau\to \mu\gamma\gamma$ decay.

\section{Final remarks}
In this work we have presented a calculation of the rare three-body FCNC top quark decay $t\to c\gamma\gamma$  in the framework of a  renormalizable model where the SM is augmented with one or three $SU(2)$ scalar LQ doublets with hypercharge $7/6$, which gives rise to two scalar LQs with electric charge of $5/3$ and $2/3$. We considered two particular scenarios: a minimal model with a lone scalar LQ doublet (scenario I) and another model with three scalar LQ doublets  $R_2^\ell$ ($\ell=\mu,\tau$) (scenario II).  While scenario I can address the muon $g-2$ anomaly, scenario II was proposed recently \cite{Crivellin:2022mff} to also explain the LFUV anomalies in $b$-hadron decays. The general aspects and the generic Lagrangian and Feynman rules for  this class of models are discussed, and analytical expressions for the one-loop LQ contribution to the invariant amplitude of  the general decay $f_i\to f_j\gamma\gamma$ are presented in terms of Passarino-Veltman integral coefficients, from which the corresponding invariant amplitude for the decay $t\to c\gamma\gamma$  follows easily.
\\

A discussion of the current bounds on the LQ masses and the LQ couplings to leptons and quarks is also presented. In scenario I the region of allowed values of LQ couplings is found by requiring  that the charge $5/3$ is responsible for  the muon $g-2$ discrepancy and obeys the constraint on the LFV decay $\tau\to \mu \gamma$ as well as the extra constraint  on the LQ couplings $|Y^{LR,RL}_{i\ell}|\le 1$, which is imposed to avoid tension with experimental data of processes sensitive to LQ contributions. It is found that the LQ couplings can be as large as $0.1$--$1$ for a LQ mass around 1 TeV, which results in a  branching ratio for the $t\to c\gamma\gamma$ decay of the order of $10^{-11}$--$10^{-12}$. When the mass of the LQ increases up to around $1.5$ TeV, this branching ratios decreases slowly but there is a high dependence on the magnitude of the LQ coupling constants.
\\

As far as scenario II is concerned, we consider the bounds obtained in the analysis of Ref. \cite{Crivellin:2022mff}, where the parameter space of the model was constrained  by requiring a solution to the LFUV anomalies and the muon $g-2$ discrepancy, along with constraints from experimental data.  For  a LQ doublet $R_2^\tau$ with a mass of $1.7$ TeV, LQ couplings to the $\tau t$ pair with values slightly larger than $O(1)$  are still allowed,  whereas the couplings of the $R_2^\mu$ doublet would be one or two orders of magnitude below for a LQ mass of 2 TeV. In this scenario the branching ratio of the $t\to c\gamma\gamma$ decay is also of the order of $10^{-11}$--$10^{-12}$.

In conclusion the LQ contribution to the branching ratio of the three body decay $t\to c\gamma\gamma$ is about two or three orders of magnitude below than  the one for the two-body decay $t\to c\gamma$, which  is below the expected experimental reach.
%As a by-product we present an estimate of the LFV decay $\tau\to \mu\gamma\gamma$, which can reach values as large as %$10^{-13}$ for a LQ with a mass of 1 TeV.

\label{conclusions}

\begin{acknowledgments}
The work of R. S\'anchez-V\'elez was partially supported by  Consejo Nacional
de Ciencia y Tecnolog\'ia under grant A1-S-23238.  G. Tavares-Velasco acknowledges partial support from Sistema Nacional de Investigadores (Mexico) and Vicerrector\'ia de Investigaci\'on y Estudios de Posgrado de la Ben\'emerita
Universidad Aut\'onoma de Puebla.

\end{acknowledgments}
\appendix

\section{Analytical results for the $f_i\to f_j\gamma\gamma$ form factors}\label{FormFactors}

The form factors ${F}_n$ $(n=1\ldots 6)$ of Eq.  \eqref{ML}  were obtained in terms of Passarino-Veltman scalar functions with the help of the FeynCalc package \cite{Mertig:1990an,Shtabovenko:2020gxv} and a cross-check was done via package-X \cite{Patel:2015tea}.  Since results in terms of Passarino-Veltman scalar functions are too lengthy to be shown here, we present our results in terms of the coefficients of two-. three-, and four-point tensor integrals: $B_1$, $B_{ii}$, $C_{i}$, $C_{ij}$, etc., where we follow the notation of Ref. \cite{Mertig:1990an}. Note however that in order to simplify our results,  a scale factor was introduced to obtain dimensionless three- and four-point scalar functions as well as dimensionless  tensor integral coefficients: all three-point scalar functions  and tensor integral coefficients, but $C_{00}(i)$, are scaled by $m_i^{-2}$, whereas all four-point scalar functions  and tensor integral coefficients, but $D_{00}(i)$ and $D_{00j}(i)$, are scaled by $m_i^{-4}$.

As already mentioned, we consider the $m_j\simeq 0$ limit (an outgoing massless fermion) and  define  Mandelstam-like scaled variables   $\hat{s}=(p'+p_2)^2/m_i^2\simeq 2p'\cdot p_2/m_i^2$, $\hat{t}=(p'+p_1)^2\simeq 2p_1\cdot p'/m_i^2$, and $\hat{u}=(p_1+p_2)^2/m_i^2=2p_1\cdot p_2/m_i^2$, which obey $\hat{s}+\hat{t}+\hat{u}=1$. We also define de auxiliary variables  $x_a=m_a^2/m_i^2$, $y_a=m_a/m_i$, for $a=k, S$, where the subscript $k$ stands for the virtual fermion. Furthermore, $\delta_{s}=\hat{s}-1$, $\delta_{t}=\hat{t}-1$, and the electric charges of the external fermions $Q_i$ and the internal one $Q_k$ are given in units of $e$.

The form factors $F_n$ ($n=1,\ldots,6$) of Eq. \eqref{ML}  can  be written as

\begin{align}
F_n
      &=\sum_{\ell_k=\mu,\tau}\left(f_n^{LL} Y^{RL}_{i\ell_k} Y^{RL}_{j\ell_k}+f_n^{RR} Y^{LR}_{i\ell_k} Y^{LR}_{j\ell_k}+f_n^{RL} Y^{LR}_{i\ell_k} Y^{RL}_{j\ell_k}
      +f_n^{LR}Y^{RL}_{i\ell_k} Y^{LR}_{j\ell_k}\right),
\end{align}
where the non-vanishing $f_n^{\ldots}$ coefficients in turn can be cast as
\begin{equation}
f_n^{\ldots}=a_n^{\ldots}\left(f_{ni}^{\ldots}Q_i^2+f_{nk}^{\ldots}Q_k^2+f_{nik}^{\ldots}Q_iQ_k\right),
\end{equation}
with $\ldots$ standing for $LL, RR$, etc. The $a_n^{\dots}$ and $f_{n\ldots}^{\ldots}$ coefficients are
\begin{equation}
a_1^{LL}=\frac{1}{\delta _s \hat{s}^2 \delta _t \hat{t}^2},
\end{equation}

\begin{align}
\label{f1LL}
f_{1i}^{LL}&=
\delta _t \left(2 \delta _s \hat{t} \left(2 \hat{s} \left(-D_{001}(2)-D_{001}(4)-2 \left(D_{003}(2)+D_{003}(4)\right)+C_2(7)\right)-2 C_{12}(4) \hat{u}+4 C_{00}(4)\right.\right.\nonumber\\&\left.\left.+B_0(3) \left(y_S^2-y_k^2-1\right)\right)+\delta _s \hat{s} \left(4 C_{00}(6)+B_0(3) \left(y_S^2-y_k^2-1\right)\right)-4 C_{12}(4) \delta _s \hat{t}^2\right)\nonumber\\&+2 \delta _s \hat{t}^2 \left(\left(B_0(1)+1\right) y_k^2-\left(B_0(2)+1\right) y_S^2-2 C_{12}(11) \hat{s}\right)+\delta _s \hat{t} \left(\hat{s} \left(4 C_{00}(11)-B_0(5)\right)\right.\nonumber\\&\left.+\left(B_0(1)+1\right) y_k^2 \left(\hat{s}-2\right)-\left(B_0(2)+1\right) \left(\hat{s}-2\right) y_S^2\right)+B_0(5) \delta _s \hat{s} \left(y_S^2-y_k^2\right),
\end{align}
\begin{align}
f_{1k}^{LL}&=\delta _t\Big(2 \delta _s \hat{s} \hat{t} \left(-\left(2 \left(D_{11}(10)+D_{11}(11)+D_{11}(12)+D_{12}(10)+D_{13}(10)+D_{13}(11)+D_{13}(12)\right)\right.\right.\nonumber\\&
      \left.\left.+2 (D_1(10)+D_1(11))+D_1(12)-D_3(12)\right) \hat{u}-2 \left(D_{00}(9)+2 D_{00}(10)+D_{00}(11)+D_{00}(12)+D_{001}(2)\right.\right.\nonumber\\&\left.\left.+D_{001}(4)+2 D_{001}(9)+2 (D_{001}(10)- D_{001}(11)- D_{001}(12))+D_{002}(9)+D_{002}(10)+2(D_{003}(2)+ D_{003}(4))\right.\right.\nonumber\\&\left.\left.+D_{003}(9)+D_{003}(10)-D_{003}(11)-D_{003}(12)\right)+C_0(13)-2 C_1(18)+2 C_2(7)+D_3(12) \delta _s\right)\nonumber\\&-2 \left(2 \left(D_{11}(10)+D_{11}(11)+D_{11}(12)+D_{12}(11)+D_{12}(12)+D_{13}(10)+D_{13}(11)+2 D_{13}(12)\right.\right.\nonumber\\&\left.\left.+D_{23}(12)+D_{33}(12)+D_1(10)+ D_1(11)+ D_1(12)\right)+D_3(12)\right) \delta _s \hat{s} \hat{t}^2\Big),
\end{align}
\begin{align}
f_{1ik}^{LL}&=\delta _t \left(2 \delta _s \hat{t} \left(\hat{s} \left(2 \left(\left(D_{11}(10)+D_{12}(10)+D_{13}(10)+D_1(10)\right) \hat{u}+D_{00}(9)+D_{002}(9)+D_{002}(10)\right.\right.\right.\right.\nonumber\\
&\left.\left.\left.\left.+2( D_{00}(10)+D_{001}(2)+ D_{001}(4)+ D_{001}(9)+ D_{001}(10))+4 (D_{003}(2)+ D_{003}(4))+D_{003}(9)+D_{003}(10)\right.\right.\right.\right.\nonumber\\
& \left.\left.\left.\left.-2 C_2(7)\right)+C_1(17)\right)+2 \left(C_{12}(3)+C_{12}(4)\right) \hat{u}-4 \left(C_{00}(3)+C_{00}(4)\right)+2 B_0(3)\right)\right.\nonumber\\&\left.+2 \delta _s \hat{s} \left(B_0(3)-2 \left(C_{00}(5)+C_{00}(6)\right)\right)+4 \delta _s \hat{t}^2 \left(C_{12}(3)+C_{12}(4)+\left(D_{11}(10)+D_{13}(10)+D_1(10)\right) \hat{s}\right)\right)
      \nonumber\\
&+2 \delta _s \hat{s} \hat{t} \left(B_0(5)-2 \left(C_{00}(10)+C_{00}(11)\right)\right)+4 \left(C_{12}(10)+C_{12}(11)\right) \delta _s \hat{s} \hat{t}^2,
\end{align}

\begin{equation}
a_1^{RL}=-\frac{2y_k}{\hat{s} \hat{t}},
\end{equation}

\begin{align}
f_{1i}^{RL}
      &=-\frac{2}{\hat{s}} \left(2 \left(D_{00}(2)+D_{00}(4)\right) \hat{s}+B_0(3) \left(\hat{s}+\hat{u}-2\right)+B_0(4)-C_0(15) \hat{s}+C_1(16) \hat{u}\right)\nonumber\\&+\frac{1}{\delta _t\delta _s \left(y_k^2-y_S^2\right)}\left(\hat{t} \left(-2 \left(\hat{s}+\hat{u}-2\right) \left(\left(B_0(1)+1\right) y_k^2-\left(B_0(2)+1\right) y_S^2\right)-2 C_2(10)\delta _s \left(y_k^2-y_S^2\right)\right)\right.\nonumber\\&\left.+\hat{t}^2 \left(2 \left(B_0(2)+1\right) y_S^2-2 \left(B_0(1)+1\right) y_k^2\right)+\left(\left(B_0(1)+1\right) y_k^2-y_S^2\right) \left(\delta _s+2 \hat{u}\right)-B_0(2) y_S^2 \left(\hat{s}+2 \hat{u}-1\right)\right)\nonumber\\&-\frac{2\hat{t}}{\hat{s}} \left(B_0(3)+C_1(16)\right) +\frac{1}{\delta_{\hat{t}} \hat{t}}\left(\delta_{\hat{t}} B_0(3)+B_0(5)\right),
\end{align}
\begin{align}
f_{1k}^{RL}&=2 (C_0(14)+ C_0(15)- D_0(7) \delta_{\hat{s}})-4 \left(D_{00}(2)+D_{00}(4)+D_{00}(9)+D_{00}(10)+D_{00}(11)+D_{00}(12)\right)\nonumber\\&+2 \left(D_1(11)+D_1(12)-D_0(5)-D_1(10)-D_2(1)+D_2(11)+D_2(12)+D_3(11)+D_3(12)\right) \hat{t}\nonumber\\&-2 \left(D_0(5)+D_1(10)-D_1(11)-D_1(12)+D_2(1)+D_2(10)-D_3(11)\right) \hat{u}+D_0(8),
\end{align}
\begin{align}
f_{1ik}^{RL}&=4 \left(2 D_{00}(2)+2 D_{00}(4)+D_{00}(9)+D_{00}(10)\right)+2 \hat{t} \left(\frac{1}{\delta _s}C_0(1)+D_0(5)+D_1(10)+D_2(1)\right)+\frac{2\hat{u}}{\delta _s} C_0(1) \nonumber\\&-\frac{1}{\delta _t}C_0(2)+\frac{1}{\hat{t}}C_0(13)-2 C_0(2)-4 C_0(15)+2 \left(D_0(5)+D_1(10)+D_2(1)+D_2(10)\right) \hat{u},
\end{align}

\begin{equation}
a_2^{RR}=-\frac{4}{\delta_{\hat{t}} \hat{s} \hat{t}^2},
\end{equation}

\begin{align}
f_{2i}^{RR}&=\delta _t \left(2 \hat{t} \left(C_{22}(6)+2 \left(D_{00}(4)+D_{001}(4)+D_{003}(2)+D_{003}(4)\right)+C_1(17)\right)+4 C_{00}(6)\right.\nonumber\\&\left.+C_{12}(6) \left(1-2 \hat{s}-2 \hat{u}\right)+B_0(3) \left(y_S^2-y_k^2-1\right)\right)+\hat{t} \left(4 C_{00}(11)+\left(B_0(1)+1\right) y_k^2-\left(B_0(2)+1\right) y_S^2\right.\nonumber\\&\left.-B_0(5)\right)+B_0(5) \left(y_S^2-y_k^2\right),
\end{align}
\begin{align}
f_{2k}^{RR}&=\delta _t \hat{t} \left(-2 \left(D_{11}(9)+D_{11}(11)+D_{11}(12)+D_{12}(9)+D_{13}(9)+D_{13}(11)\right.\right.\nonumber\\&\left.\left.+D_{13}(12)+D_1(9)+D_1(11)+D_1(12)\right) \hat{s}+4 \left(D_{00}(4)+ D_{00}(9)- D_{00}(12)\right)+D_{11}(9)+D_{11}(11)+D_{11}(12)\right.\nonumber\\&\left.+D_{12}(9)+D_{13}(9)+D_{13}(11)+D_{13}(12)+4 \left(D_{001}(4)+D_{001}(9)+D_{001}(10)-D_{001}(11)-D_{001}(12)+D_{002}(9)\right.\right.\nonumber\\&\left.\left.+D_{003}(2)+D_{003}(4)+D_{003}(9)-D_{003}(12)\right)+2 \left(D_0(6)+D_1(3)+D_1(9)+D_2(3)+D_2(9)\right) y_k^2\right.\nonumber\\&\left.+D_1(9)+D_1(11)+D_1(12)\right),
\end{align}
\begin{align}
f_{2ik}^{RR}&=\delta _t \left(\hat{t} \left(-2 (C_{22}(5)+C_{22}(6))+2 \left(D_{11}(9)+D_{12}(9)+D_{13}(9)+D_1(9)\right) \hat{s}\right.\right.\nonumber\\&\left.\left.-4(2 D_{00}(4)+ D_{00}(9))-D_{11}(9)-D_{12}(9)-D_{13}(9)-4 \left(2 D_{001}(4)+D_{001}(9)+D_{001}(10)+D_{002}(9)\right.\right.\right.\nonumber\\&\left.\left.\left.+2 \left(D_{003}(2)+D_{003}(4)\right)+D_{003}(9)\right)-4 C_1(17)-2 \left(D_0(6)+D_1(3)+D_1(9)+D_2(3)+D_2(9)\right) y_k^2-D_1(9)\right)\right.\nonumber\\&\left.+\left(C_{12}(5)+C_{12}(6)\right) \left(2 \hat{s}+2 \hat{u}-1\right)+2 B_0(3)\right)+2 \hat{t} \left(B_0(5)-2 \left(C_{00}(5)+C_{00}(6)+C_{00}(10)+C_{00}(11)\right)\right)\nonumber\\&+4 \left(C_{00}(5)+C_{00}(6)\right),
\end{align}

\begin{equation}
a_2^{LR}=\frac{4y_k}{\hat{s} \hat{t} },
\end{equation}

\begin{align}
f_{2i}^{LR}&=\frac{1}{\hat{t}}\left( B_0(3)-C_1(17)\right)+\frac{1}{\delta _t}\left(\frac{2}{y_k^2-y_S^2}\left( \left(B_0(2)+1\right) y_S^2-\left(B_0(1)+1\right) y_k^2\right)+\frac{2}{\hat{t}} B_0(5)\right),
\end{align}
\begin{align}
f_{2k}^{LR}&=2 D_0(8)-D_0(6)-D_1(9)+D_1(11)+D_1(12)-D_2(3)-D_2(9)+D_3(12),
\end{align}
\begin{align}
f_{2ki}^{LR}&=\frac{2}{\delta _t} C_0(2)+\frac{1}{\hat{t}}C_0(13)+D_0(6)+D_1(9)+D_2(3)+D_2(9),
\end{align}
\begin{equation}
a_3^{LL}=-\frac{4}{\hat{s} \hat{t}},
\end{equation}
\begin{align}
f_{3i}^{LL}&=D_{13}(2)+D_{13}(4)+2 (D_{33}(2)+D_{33}(4)+D_{333}(2)+D_{333}(4))+D_{113}(2)+D_{113}(4)+3 (D_{133}(2)+ D_{133}(4)),
\end{align}
\begin{align}
f_{3k}^{LL}&=2 (D_{11}(9)+ D_{11}(10))-4( D_{11}(11)- D_{11}(12))+D_{12}(9)+D_{12}(10)+D_{13}(2)\nonumber\\&+D_{13}(4)+D_{13}(9)+D_{13}(10)+2 (D_{33}(2)+D_{33}(4)+ D_{111}(9)+ D_{111}(10)- D_{111}(11)- D_{111}(12))\nonumber\\&-3 (D_{13}(11)+ D_{13}(12)-D_{112}(9)-D_{112}(10)-D_{113}(9)-D_{113}(10)+D_{113}(11)+ D_{113}(12)\nonumber\\&-D_{133}(2)- D_{133}(4))+D_{113}(2)+D_{113}(4)+D_{122}(9)+D_{122}(10)+2 (D_{123}(9)+ D_{123}(10))+D_{133}(9)\nonumber\\&+D_{133}(10)-D_{133}(11)-D_{133}(12)+2 \left(D_{333}(2)+D_{333}(4)\right)-2 D_1(11)-2 D_1(12),
\end{align}
\begin{align}
f_{3ik}^{LL}&=-\Big(2 (D_{11}(9)+ D_{11}(10))+D_{12}(9)+D_{12}(10)+2 (D_{13}(2)+ D_{13}(4))+D_{13}(9)+D_{13}(10)+4(D_{33}(2)+ D_{33}(4))
      \nonumber\\&+2( D_{111}(9)+ D_{111}(10))+3( D_{112}(9)+ D_{112}(10))+2 (D_{113}(2)+ D_{113}(4))+3 (D_{113}(9)+ D_{113}(10))\nonumber\\&+D_{122}(9)+D_{122}(10)+2(D_{123}(9)+ D_{123}(10))+6 (D_{133}(2)+ D_{133}(4))+D_{133}(9)+D_{133}(10)\nonumber\\&+4 \left(D_{333}(2)+D_{333}(4)\right),
\end{align}

\begin{equation}
a_3^{RL}=\frac{8y_k}{\hat{s} \hat{t}},
\end{equation}
\begin{align}
f_{3i}^{RL}&=D_{13}(2)+D_{13}(4)+D_{33}(2)+D_{33}(4)+D_3(2)+D_3(4),
\end{align}
\begin{align}
f_{3ik}^{RL}&=D_{11}(9)+D_{11}(10)
      +D_{11}(11)+D_{11}(12)+D_{12}(9)+D_{12}(10)+D_{13}(2)+D_{13}(4)+D_{13}(9)+D_{13}(10)
      \nonumber\\&+D_{13}(11)+D_{13}(12)+D_{33}(2)
      +D_{33}(4)+D_1(9)+D_1(10)+D_1(11)+D_1(12)+D_3(2)+D_3(4),
\end{align}
\begin{align}
f_{3k}^{RL}&=
      -\Big(D_{11}(9)+D_{11}(10)+D_{12}(9)+D_{12}(10)+2 (D_{13}(2)+ D_{13}(4))+D_{13}(9)+D_{13}(10)+2 \left(D_{33}(2)+D_{33}(4)\right)\nonumber\\&+D_1(9)+D_1(10)+2 (D_3(2)+ D_3(4))\Big),
\end{align}
\begin{equation}
a_4^{RR}=\frac{1}{\hat{s} \delta_{\hat{t}} \hat{t}^2},
\end{equation}
\begin{align}
f_{4i}^{RR}&=\delta _t \left(4 C_{00}(6)+4 \left(D_{001}(2)-D_{001}(4)\right) \hat{t}+B_0(3) \left(y_S^2-y_k^2-1\right)\right)\nonumber\\&+\hat{t} \left(4 C_{00}(11)+\left(B_0(1)+1\right) y_k^2-\left(B_0(2)+1\right) y_S^2-B_0(5)\right)+B_0(5) \left(y_S^2-y_k^2\right),
\end{align}
\begin{align}
f_{4k}^{RR}&=2 \delta _t \hat{t} \left(-2 \left(D_{00}(9)+D_{00}(11)+D_{00}(12)-D_{001}(2)+D_{001}(4)+D_{002}(9)-D_{002}(10)+D_{003}(9)-D_{003}(10)
      \right.\right.\nonumber\\&\left.\left.+D_{003}(11)-D_{003}(12)\right)+C_0(13)+D_1(12) \hat{u}\right),
\end{align}
\begin{align}
f_{4ik}^{RR}&=2 \hat{t} \left(B_0(5)-2 \left(C_{00}(10)+C_{00}(11)\right)\right)+\delta _t \left(2 \left(B_0(3)-2 \left(C_{00}(5)+C_{00}(6)\right)\right) \right.\nonumber\\&\left.+2 \hat{t} \left(2 \left(D_{00}(9)-2 D_{001}(2)+2 D_{001}(4)+D_{002}(9)-D_{002}(10)+D_{003}(9)-D_{003}(10)\right)+C_1(17)\right)\right),
\end{align}
\begin{equation}
a_4^{LR}=\frac{2y_k}{\hat{s} \delta_{\hat{t}} \hat{t}^2},
\end{equation}

\begin{align}
f_{4i}^{LR}&=\frac{\hat{t}}{y_k^2-y_S^2} \left(\left(B_0(1)+1\right) y_k^2-\left(B_0(2)+1\right) y_S^2\right)-B_0(5)-B_0(3)\delta_{\hat{t}},
\end{align}
\begin{align}
f_{4k}^{LR}&=-D_0(8) \hat{t}\delta_{\hat{t}},
\end{align}
\begin{align}
f_{4ik}^{LR}&=-\left(C_0(2) \hat{t}+C_0(13)\delta_{\hat{t}}\right),
\end{align}
\begin{equation}
a_5^{LL}=\frac{4}{\hat{s} \hat{t}^2},
\end{equation}
\begin{align}
f_{5i}^{LL}&=C_{12}(6),
\end{align}
\begin{align}
f_{5k}^{LL}&= \hat{t}\left(D_{11}(9)+D_{11}(11)+D_{11}(12)+D_{12}(9)+D_{13}(9)+D_{13}(11)+D_{13}(12)+D_1(9)+D_1(11)+D_1(12)\right),
\end{align}
\begin{align}
f_{5ik}^{LL}&=-\left(C_{12}(5)+C_{12}(6)+\hat{t}\left(D_{11}(9)+D_{12}(9)+D_{13}(9)+D_1(9)\right)\right),
\end{align}
\begin{equation}
a_5^{RL}=-\frac{4y_k}{\hat{s} \hat{t}^2},
\end{equation}

\begin{align}
f_{5i}^{RL}&=C_1(17),
\end{align}
\begin{align}
f_{5k}^{RL}&=\hat{t}\left(D_0(6)+D_1(9)-D_1(11)-D_1(12)+D_2(3)+D_2(9)-D_3(12)\right),
\end{align}
\begin{align}
f_{5ik}^{RL}&=C_0(13)-\hat{t}\left(D_0(6)+D_1(9)+D_2(3)+D_2(9)\right),
\end{align}
and
\begin{equation}
a_6^{RR}=\frac{4}{\hat{s} \hat{t}},
\end{equation}
\begin{align}
\label{f6RR}
f_{6i}^{RR}&=D_{13}(2)-D_{13}(4)+D_{113}(2)-D_{113}(4)+D_{133}(2)-D_{133}(4),
\end{align}
\begin{align}
f_{6k}^{RR}&=
-D_{12}(9)+D_{12}(10)+D_{13}(2)-D_{13}(4)-D_{13}(9)+D_{13}(10)-D_{13}(11)+D_{13}(12)-D_{112}(9)+D_{112}(10)\nonumber\\&
      +D_{113}(2)-D_{113}(4)-D_{113}(9)+D_{113}(10)-D_{113}(11)+D_{113}(12)-D_{122}(9)+D_{122}(10)-2 D_{123}(9)\nonumber\\&+2 D_{123}(10)+D_{133}(2)-D_{133}(4)-D_{133}(9)+D_{133}(10)-D_{133}(11)+D_{133}(12),
\end{align}
\begin{align}
f_{6ik}^{RR}=&D_{12}(9)-D_{12}(10)-2 D_{13}(2)+2 D_{13}(4)+D_{13}(9)-D_{13}(10)+D_{112}(9)-D_{112}(10)-2 D_{113}(2)\nonumber\\&+2 D_{113}(4)+D_{113}(9)-D_{113}(10)+D_{122}(9)-D_{122}(10)+2 D_{123}(9)-2 D_{123}(10)-2 D_{133}(2)+2 D_{133}(4)\nonumber\\&+D_{133}(9)-D_{133}(10).
\end{align}

As far as the arguments of the Passarino-Veltman scalar functions scalar functions and tensor integral coefficients are concerned, they are presented in Tables \ref{B0arguments} through \ref{Darguments}, where again we follow the notation of \cite{Mertig:1990an}.

\begin{table}
\caption{Arguments of the two-point scalar functions $B_0(i)$ in the notation of \cite{Mertig:1990an}.\label{B0arguments}}
\renewcommand{\arraystretch}{1.4}
\begin{tabular}{>{$}c<{$} >{$}c<{$}}
\hline
(i)&(a,b,c)\\
\hline
 (1) & \left(0,m_k^2,m_k^2\right) \\
 (2) & \left(0,m_S^2,m_S^2\right) \\
 (3) & \left(m_i^2,m_k^2,m_S^2\right) \\
 (4) & \left(s,m_k^2,m_S^2\right) \\
 (5) & \left(t,m_k^2,m_S^2\right) \\
 \hline
\end{tabular}
\end{table}

\begin{table}
\caption{Arguments of the three-point scalar functions $C_0(i)$ and three-point tensor integral coefficients $C_j(i)$ and $C_{jk}(i)$ in the notation of \cite{Mertig:1990an}.\label{Carguments}}
\renewcommand{\arraystretch}{1.4}
\begin{tabular}{>{$}c<{$} >{$}c<{$}}
\hline
(i)&(a,b,c,d,e,f,g)\\
\hline
 (1) & \left(0,0,s,m_k^2,m_k^2,m_S^2\right) \\
 (2) & \left(0,0,t,m_k^2,m_k^2,m_S^2\right) \\
 (3) & \left(0,m_i^2,s,m_k^2,m_k^2,m_S^2\right) \\
 (4) & \left(0,m_i^2,s,m_S^2,m_S^2,m_k^2\right) \\
 (5) & \left(0,m_i^2,t,m_k^2,m_k^2,m_S^2\right) \\
 (6) & \left(0,m_i^2,t,m_S^2,m_S^2,m_k^2\right) \\
 (7) & \left(0,m_i^2,u,m_S^2,m_k^2,m_S^2\right) \\
 (8) & \left(0,s,0,m_k^2,m_k^2,m_S^2\right) \\
 (9) & \left(0,s,0,m_S^2,m_S^2,m_k^2\right) \\
 (10) & \left(0,t,0,m_k^2,m_k^2,m_S^2\right) \\
 (11) & \left(0,t,0,m_S^2,m_S^2,m_k^2\right) \\
 (12) & \left(m_i^2,0,s,m_S^2,m_k^2,m_k^2\right) \\
 (13) & \left(m_i^2,0,t,m_S^2,m_k^2,m_k^2\right) \\
 (14) & \left(m_i^2,0,u,m_k^2,m_S^2,m_k^2\right) \\
 (15) & \left(m_i^2,0,u,m_S^2,m_k^2,m_S^2\right) \\
 (16) & \left(m_i^2,s,0,m_k^2,m_S^2,m_k^2\right) \\
 (17) & \left(m_i^2,t,0,m_k^2,m_S^2,m_k^2\right) \\
 (18) & \left(m_i^2,u,0,m_S^2,m_k^2,m_k^2\right) \\
 \hline
\end{tabular}
\end{table}

\begin{table}
\caption{Arguments of the four-point scalar functions $D_0(i)$ and four-point tensor integral coefficients $D_j(i)$, $D_{jk}(i)$ and $D_{jkl}(i)$ in the notation of \cite{Mertig:1990an}. \label{Darguments}}
\renewcommand{\arraystretch}{1.4}
\begin{tabular}{>{$}c<{$} >{$}c<{$}}
\hline
(i)&(a,b,c,d,e,f,g,h,i,j)\\
\hline
 (1) & \left(0,s,0,t,0,m_i^2,m_k^2,m_k^2,m_S^2,m_S^2\right) \\
 (2) & \left(0,s,m_i^2,u,0,0,m_S^2,m_S^2,m_k^2,m_S^2\right) \\
 (3) & \left(0,t,0,s,0,m_i^2,m_k^2,m_k^2,m_S^2,m_S^2\right) \\
 (4) & \left(0,t,m_i^2,u,0,0,m_S^2,m_S^2,m_k^2,m_S^2\right) \\
 (5) & \left(m_i^2,0,0,0,s,t,m_k^2,m_S^2,m_S^2,m_k^2\right) \\
 (6) & \left(m_i^2,0,0,0,s,t,m_S^2,m_k^2,m_k^2,m_S^2\right) \\
 (7) & \left(m_i^2,0,0,0,s,u,m_S^2,m_k^2,m_k^2,m_k^2\right) \\
 (8) & \left(m_i^2,0,0,0,t,u,m_S^2,m_k^2,m_k^2,m_k^2\right) \\
 (9) & \left(m_i^2,s,0,t,0,0,m_k^2,m_S^2,m_k^2,m_S^2\right) \\
 (10) & \left(m_i^2,t,0,s,0,0,m_k^2,m_S^2,m_k^2,m_S^2\right) \\
 (11) & \left(m_i^2,u,0,s,0,0,m_S^2,m_k^2,m_k^2,m_k^2\right) \\
 (12) & \left(m_i^2,u,0,t,0,0,m_S^2,m_k^2,m_k^2,m_k^2\right) \\
 \hline
\end{tabular}
\end{table}

\section{Average square amplitude for the $f_i\to f_j\gamma\gamma$ decay}
After averaging (summing) over polarizations of the ingoing fermion (outgoing particles), from Eq. \eqref{ML} we obtain the following average square amplitude in the $m_j\simeq 0$ limit
\label{squareampli}
\begin{equation}
\label{sqramp}
\left|\overline{\mathcal{M}}(f_i\to f_j\gamma\gamma)\right|^2
=\frac{N_c^2 \alpha^2}{16\pi^2}\left({\cal A}_{1}+{\cal A}_2\right),
\end{equation}
with
\begin{align*}
{\cal A}_1&=\Bigg\{\frac{\hat{s} \hat{t}}{32}\Big[\left(\hat{s}+\hat{t}\right)\left(4 \hat{s} \lvert| F_1\rvert| {}^2+2 \hat{t} \hat{u}\lvert| F_2\rvert| {}^2+\hat{s}  \hat{u}^2\lvert| F_3\rvert| {}^2\right)\nonumber\\&+ \left(\left(\hat{s}-\hat{t}\right){}^2
+\hat{u}(\hat{s}+\hat{t})\right)\left(4 \hat{s}\lvert| F_4\rvert| {}^2+2\hat{t} \hat{u}\lvert| F_5\rvert| {}^2
+\hat{s} \hat{u}^2 \lvert| F_6\rvert| {}^2\right)
+2 \hat{s}  \left(\hat{s}+\hat{t}\right) \hat{u}{\rm Re}\left(F_1 F_3^*\right)
\nonumber\\&
+\hat{u} \left(\left(\hat{s}-\hat{t}\right){}^2-(\hat{s}+\hat{t})^2\right){\rm Re}\Big(
\left(2F_1+\hat{u}F_3\right) F_5^*-F_2\left(2F_4^*+\hat{u} F_6^*\right)\Big)
+2 \hat{s} \hat{u} \left(\left(\hat{s}-\hat{t}\right){}^2+\hat{u}(\hat{s}+\hat{t})\right){\rm Re}\left(F_4 F_6^*\right)\nonumber\\
&+ 2\hat{t} \left(\hat{t}-\hat{s}\right){\rm Re}\left(\hat{s}(4F_1+\hat{u}F_3) F_4^*+2 \hat{t} \hat{u}F_2 F_5^*+\hat{s} \hat{u}(F_1+\hat{u}F_3) F_6^*\right)\Big]+(\hat{s}\leftrightarrow \hat{t})\Bigg\}+
(Y_{lk}^{LR}\leftrightarrow Y_{lk}^{RL}),
\end{align*}

and
\begin{align}
{\cal A}_2=&\frac{1}{32} \hat{s} \hat{t}^2{\rm Re}\Big[2 \hat{s} \left(\hat{s}+\hat{t}\right) \hat{u}\left(F_3 G_1^*-F_2 G_2^*+F_1 G_3^*+
\hat{u}F_3 G_3^*\right)\nonumber\\&+
\hat{u}^2 \left( \left(\hat{s}-\hat{t}\right)^2-\left(\hat{s}+\hat{t}\right)^2\right)\left(\frac{\hat{s}}{\hat{t}}
\left(F_6 G_2^*+F_3 G_5^*\right)+F_5 G_3^*+F_2 G_6^*
\right)\nonumber\\&
+2 \hat{s} \hat{u} \left(\left(\hat{s}-\hat{t}\right)^2+\hat{u}\left(\hat{s}+\hat{t}\right)\right)
\left(F_5 G_5^*-F_6 G_6^*-F_6 G_4^*-F_4 G_6^*\right)\nonumber\\
&+2\hat{s} \hat{u} \left(\hat{t}-\hat{s}\right)\left(F_6 G_1^*-F_1 G_6^*+F_2 G_5^*-F_5 G_2^*+F_4 G_3^*-F_3 G_4^*+\hat{u}F_6 G_3^*-\hat{u}F_3 G_6^*\right)\Big]+(\hat{s}\leftrightarrow \hat{t}),
\end{align}
where $G_n(\hat{s},\hat{t})=F_n(\hat{t},\hat{s})$ are the form factors obtained after exchanging the photons. The extra terms arise from the right-handed amplitude ${\cal M}_R$ and interference terms.

\section{LQ contribution to lepton process}
\label{LFVformulas}For completeness we present some formulas used in the evaluation  of the constraints on the LQ couplings presented in Sec. \ref{LQcoupcons}.
\subsection{Muon anomalous magnetic dipole moment}

The one-loop   contribution   to the muon anomalous magnetic dipole moment $a_{\mu}^{\rm LQ}$ from  a charge $Q_S$ scalar LQ with couplings  to the muon and quark $q_i$
 can be written as \cite{Bolanos:2013tda}

\begin{align}
\label{afi}
a_{\mu}^{\rm LQ}&=-\sum_ {i}\frac{3\,\sqrt{x_{\mu}}}{32\pi^2} \Bigg(\sqrt{x_\mu}\left(\left|Y_{i\mu}^{RL}\right|^2+
\left|Y_{i\mu}^{LR}\right|^2\right) F\left(x_\mu,x_{q_i}\right) +2\,\sqrt{x_{q_i}}\,{\rm Re}\left(Y_{i\mu}^{RL}
{Y_{i\mu}^{LR}}^*\right) G\left(x_\mu,x_{q_i}\right) \Bigg),
\end{align}
where $x_a=m_a^2/m^2_{S}$ and the sum runs over the second and third generation up  (down) quarks for $Q_S=5/3$ ($Q_S=2/3$). The  $F(x,y)$ and $G(x,y)$ functions are given in terms of Feynman parameter integrals  by
\begin{eqnarray}
\label{Fagfeyn}
F(z_1,z_2)=2\int_0^1\frac{(1-x) x \left(Q_{q_i}(1-x)+Q_S x\right)}{(1-x)(z_2- x z_1)+x}dx,\\
\label{Gagfeyn}
G(z_1,z_2)=2\int_0^1\frac{(1-x) \left(Q_{q_i}(1-x)+Q_S x\right)}{(1-x)(z_2- x z_1)+x}dx,
\end{eqnarray}
where the electric charges of the internal quark and the LQ $q_{i}$ and $Q_S$ are in units of the positron charge. In the limit of a heavy quark, $a_\mu^{\rm LQ}$  reduces to
\begin{align}
a_{\mu}^{\rm LQ}&\simeq\frac{3}{16\pi^2}\sum_ {i}\frac{\sqrt{x_{\mu}}\sqrt{x_{q_i}}}{(1-x_{q_i})^3}\,{\rm Re}\left(Y_{i\mu}^{RL}
Y_{i\mu}^{LR*}\right) \left(Q_{q_i}\left(3-4x_{q_i}+x_{q_i}^2+2 \log (x_{q_i})\right)\right.\nonumber\\
&\left.-Q_S\left(1-x_{q_i}^2+2 x_{q_i}
\log (x_{q_i})\right)\right).
\end{align}
We note that since $\Omega_{2/3}$  is a chiral LQ (it only has either left- or right-handed couplings to fermion pairs),  its contribution to $a_\mu$ can be neglected as it is proportional to  $x_\mu$ rather than $\sqrt{x_{\mu}}\sqrt{x_{q_i}}$.

\subsection{$\tau\to \mu \gamma$ decay}
A charge $Q_S$ scalar LQ contribute at the one-loop level to the $\tau\to \mu\gamma$ decay via triangle diagrams with  an internal LQ. The corresponding decay width   is given by
\begin{align}
\Gamma({\tau}\to {\mu} \gamma)&=\frac{ m_{\tau}}{16 \pi   } \left(1-\left(\frac{m_{\mu}}{m_{\tau}}\right)^2\right)^3\left(|L|^2+|R|^2\right),
\end{align}
where the $L$ and $R$ form factors are ultraviolet finite and are given in terms of Feynman parameter integrals as follows \cite{Bolanos:2013tda}
\begin{equation}
L=\frac{3g^2e\sqrt{x_{\tau}}}{64c_W^2\pi^2}\sum_{i}\left(\sqrt{x_{\tau}} Y^{RL}_{i\tau} Y^{RL}_{i\mu}H_1(x_{q_i})+
\sqrt{x_{\mu}} Y^{LR}_{i\tau} Y^{LR}_{i\mu}H_2(x_{q_i})+
\sqrt{x_{q_i}}Y^{RL}_{i\mu}Y^{LR}_{i\tau} H_3(x_{q_i})\right),
\end{equation}
where
\begin{align}
H_1(z)&=\int_0^1 dx \int_0^{1-x}dyx\left(\frac{Q_{q_i} y}{\zeta _1(x_\tau,x_\mu,z)}
-\frac{Q_S(1-x-y)}{\zeta _2(x_\tau,x_\mu,z)} \right),
\end{align}
\begin{align}
H_2(z)&=\int_0^1 dx \int_0^{1-x}dy(1-x-y)\left(\frac{Q_{q_i}x}{\zeta _1(x_\tau,x_\mu,z)} -
\frac{Q_S y}{\zeta _2(x_\tau,x_\mu,z)}\right),
\end{align}
and
\begin{align}
H_3(z)&=\int_0^1 dx \int_0^{1-x}dy\left(\frac{Q_{q_i}(1-x)}{\zeta _1(x_\tau,x_\mu,z)}
-\frac{Q_S(1-x-y)}{\zeta _2(x_\tau,x_\mu,z)}  \right),
\end{align}
with
\begin{align}
\zeta_1(z_1,z_2,z_3)&=x \left(y(z_1-z_2)+z_2 (x-1)+1-z_3\right)+z_3,\\
\zeta_2(z_1,z_2,z_3)&=x y \left(z_2-z_1\right)-x \left((1-x)z_1-z_3-1\right)-y \left((1-y)z_2-z_3-1\right)+z_3.
\end{align}
In addition, the right handed form factor can be obtained from the left-handed one as follows
\begin{equation}
\label{Rgamma}
R=L\left(\begin{array}{c}Y_{m\ell_k}^{RL}\leftrightarrow Y_{m\ell_k}^{LR}\\ Q_S\to -Q_S\end{array}\right).
\end{equation}
We note that the results presented in this Appendix  are given in terms of Passarino-Veltman scalar functions in \cite{Bolanos:2013tda}.

\bibliography{Biblio}
\end{document}